\documentclass{aa}  

\usepackage{graphicx}

\usepackage{txfonts}

\usepackage{booktabs}
\usepackage{xcolor}
\usepackage{pdflscape}
\usepackage{hyperref}

\begin{document} 

   \title{CHEX-MATE: exploring the kinematical properties of Planck galaxy clusters
   \thanks{Based in part on observations collected at the European Southern Observatory under ESO programmes 0110.A-4192 and 0111.1-0186.}}

%   \subtitle{CHEX-MATE: kinematical and dynamical properties}

   \author{Lorenzo Pizzuti
          \inst{\ref{unimib}}
          \and
          Rafael Barrena\inst{\ref{iac},\ref{ull}}
          \and
          Mauro Sereno\inst{\ref{ossbo},\ref{infn_inaf}}
          \and
          Alina Streblyanska\inst{\ref{iac},\ref{ull}}
          \and 
          Antonio Ferragamo\inst{\ref{uninapo}}
          \and
          Sophie Maurogordato\inst{\ref{obs_cot_azur}}
          \and 
          Alberto Cappi\inst{\ref{ossbo},\ref{obs_cot_azur}}
          \and
          Stefano Ettori\inst{\ref{ossbo},\ref{infn_inaf}}
          \and
          Gabriel W. Pratt\inst{\ref{saclay}}
          \and
        Gianluca Castignani\inst{\ref{ossbo}}
         \and
        Megan Donahue\inst{\ref{michi}}
        \and
        Dominique Eckert\inst{\ref{ossbo},\ref{infn_inaf}}
        \and
        Fabio Gastaldello \inst{\ref{IASF}}
        \and
        Raphael Gavazzi\inst{\ref{marseille},\ref{iap}}
        \and
        Christopher P. Haines\inst{\ref{atacama}}
        \and
        Scott T. Kay\inst{\ref{jordell}}
        \and
        Lorenzo Lovisari\inst{\ref{IASF},\ref{harv}}
        \and
        Ben J. Maughan \inst{\ref{Bristol}}
        \and 
        Etienne Pointecouteau \inst{\ref{touluse}}
        \and
        Elena Rasia \inst{\ref{oats},\ref{ifpu},\ref{unimic}}
        \and
        Mario Radovich\inst{\ref{oap}}
        \and
        Jack Sayers \inst{\ref{caltech}}
          }       
   \institute{
   Dipartimento di Fisica G. Occhialini, Universit\`a degli Studi di Milano Bicocca, Piazza della Scienza 3, I-20126 Milano, Italy\label{unimib}
   \and
   Instituto de Astrof\'{\i}sica de Canarias, C/ V\'{\i}a L\'{a}ctea s/n, E-38205 La Laguna, Tenerife, Spain \label{iac}
   \and
   Universidad de La Laguna, Departamento de Astrof\'{i}sica, E-38206 La Laguna, Tenerife, Spain \label{ull}
   \and
   INAF \text{--} Osservatorio di Astrofisica e Scienza dello Spazio di Bologna, via Piero Gobetti 93/3, I-40129 Bologna, Italy \label{ossbo}
   \and 
   INFN, Sezione di Bologna, viale Berti Pichat 6/2, I-40127 Bologna, Italy \label{infn_inaf}
   \and
   Dipartimento di Fisica 'E. Pancini', Universit\`a degli Studi di Napoli Federico II, Via Cinthia, 21, I-80126 Napoli, Italy \label{uninapo}
   \and
   Universit\'e C$\hat{\textrm{o}}$te d'Azur, Observatoire de la C$\hat{\textrm{o}}$te d'Azur, CNRS, Laboratoire Lagrange, Bd de l'Observatoire, CS 34229, 06304 Nice Cedex 4, France \label{obs_cot_azur}
   \and
   Universit\'e Paris-Saclay, Universit\'e Paris Cit\'e, CEA, CNRS, AIM, 91191 Gif-sur-Yvette, France \label{saclay}
   \and
   Michigan State University, Physics and Astronomy Department, 567 Wilson Road, East Lansing, Michigan 48824, USA \label{michi}
   \and
   INAF, Istituto di Astrofisica Spaziale e Fisica Cosmica di Milano, via A. Corti 12, 20133 Milano, Italy \label{IASF}
   \and
   Laboratoire d’Astrophysique de Marseille, Aix-Marseille Univ., CNRS, CNES, 13013 Marseille, France  \label{marseille}
   \and Institut d’Astrophysique de Paris, UMR7095 CNRS \& Sorbonne Universite\', 98bis Bd Arago, 75014, Paris, France \label{iap}
      \and
    Instituto de Astronomía y Ciencias Planetarias, Universidad de Atacama, Copayapu 485, Copiapó, Chile\label{atacama}
    \and
   Jodrell Bank Centre for Astrophysics, Department of Physics and Astronomy, University of Manchester, Manchester M13 9PL, UK \label{jordell}
   \and 
   Center for Astrophysics $|$ Harvard $\&$ Smithsonian, 60 Garden Street, Cambridge, MA 02138, USA \label{harv}
   \and 
   HH Wills Physics Laboratory, University of Bristol, Tyndall Ave, Bristol, BS8 1TL, UK\label{Bristol}
   \and
   IRAP, CNRS, Université de Toulouse, CNES, Toulouse, France \label{touluse}
   \and
   Osservatorio Astronomico di Trieste, via Tiepolo 11, I-34131, Trieste, Italy\label{oats}
   \and
   IFPU, Institute for Fundamental Physics of the Universe, Via Beirut 2, 34014 Trieste, Italy\label{ifpu}
   \and
   Department of Physics; University of Michigan, Ann Arbor, MI 48109, USA\label{unimic}
   \and
   INAF - Osservatorio Astronomico di Padova, vicolo Osservatorio 5, I-35122, Padova, Italy\label{oap}
   \and
    California Institute of Technology, Pasadena, CA 91125 USA\label{caltech}
   }
%             \email{lorenzo.pizutti@inaf.it}
%             \thanks{bla, bla, bla ...}

   \date{Received XXX; accepted XXX}

% \abstract{}{}{}{}{} 
% 5 {} token are mandatory

%The characterisation of dynamical state of galaxy clusters in large surveys is essential for making unbiased use of cosmological samples. In this context, w 
  \abstract{We analyse the kinematical properties of the CHEX-MATE (Cluster HEritage project with XMM-Newton – Mass Assembly and Thermodynamics at the Endpoint of structure formation) galaxy cluster sample. Our study is based on the radial velocities retrieved from SDSS-DR18, DESI and NED spectroscopic databases and new data obtained with the 10.4m GTC or the ESO-NTT telescopes. We derive cluster mass profiles for 75 clusters using the \textsc{MG-MAMPOSSt} procedure, which recovers the gravitational potential and the anisotropy profiles from line-of-sight velocities and projected positions of galaxy members. The standard NFW and the Burkert models with flatter cores than NFW both adequately fit the kinematic data, with only marginal statistical preference for one model over the other. An estimation of the mass bias $(1-B_1) = M^{SZ}_{500}/M^{M}_{500}$ is performed from the comparison with SZ-X-ray-calibrated mass estimates, resulting in a value of $0.54 \pm 0.11$ when four evidently disturbed clusters are removed from the sample. 
  We assess the dynamical state of the clusters by inferring the Anderson-Darling coefficient ($A^2$) and the fraction of galaxies in substructures ($f_\text{sub}$). Except for a few cases, we found relatively low values for $A^2$, suggesting that CHEX-MATE clusters are not too far from relaxation. Moreover, no significant trends emerge among $A^2,\,f_\text{sub}$ and the difference between the log-masses estimated by \textsc{MG-MAMPOSSt} and by SZ-X-ray. 
  We study the concentration-mass relation for the sample; despite the large scatter, we observe signs of an increasing trend for large-mass clusters, in agreement with recent theoretical expectations.
  
  Finally, the analysis of radial anisotropy profiles of member galaxies - stacked in five bins of mass and redshift - reveals that orbits tend to be isotropic at the center and more radial towards the edge, as already found in previous studies. A slight trend of increasing radial orbits at $r_{200}$ is observed in clusters with larger velocity dispersion.}


   \maketitle
%
%________________________________________________________________

\section{Introduction}
\label{sec:intro}
Galaxy clusters are the most massive virialised structures in the Universe, so they are excellent natural laboratories to study the formation, structure assembly, and evolution of mass halos and subhalos across the cosmic time. The combination of optical data, X-ray observations, and the signal from Sunyaev-Zel'dovich (SZ) Effect \citep{SZ72} provides the ideal scenario to study galaxy clusters. A multi-wavelength analysis will allow us to acquire a complete panorama from the physics behind different components: Dark Matter (DM; \citealt{Zw33}), gas \citep{Cav81}, and galaxies (e.g., \citealt{Abell58}), and see how they interact to configure these such massive objects.

In this work we study dynamical cluster mass profiles, the presence of substructures, and velocity anisotropies by analysing the projected phase-space distribution of member galaxies (e.g., \citealt{Carl97}) under the assumption of dynamical equilibrium. In fact, as galaxies enter into clusters through hierarchical accretion, their orbits contain important information on the processes that lead to the cluster mass assembly and internal dynamical relaxation (e.g., \citealt{Biv_Kat04}). Moreover, knowledge of the orbits of cluster galaxies is crucial to understand environmental effects that differentiate galaxy evolution in clusters relative to the field (\citealt{Lotz19}; \citealt{Tonn19}).

Our studies are carried out using the \textsc{MG-MAMPOSSt} (Modelling Anisotropy and Mass Profile of Spherical Observed Systems) code of \cite{Pizzuti:2022ynt}. This technique allows us to obtain a cluster mass profile reconstruction and the velocity anisotropy profile for several parametrisations of the total gravitational potential, in GR and in general dark energy or modified gravity frameworks.  In other words, \textsc{MG-MAMPOSSt} jointly obtains the best fit parameters for the cluster mass distribution and for the velocity anisotropy profile of cluster members (\citealt{Mun14}; \citealt{Zara21}); this analysis allows us to retrieve information about orbits of galaxies, by distinguishing, statistically, between populations with radial and/or circular configurations.

We developed our analysis in the CHEX-MATE cluster sample \citep{chex-mate21}, the 
Cluster HEritage project with XMM-Newton – Mass Assembly and Thermodynamics at the Endpoint of structure formation. This project is based on the three-mega-second Multi-Year Heritage Programme to obtain X-ray observations of a minimally-biased, signal-to-noise-limited (SNR$>6.5$) sample of 118 galaxy clusters detected by {\it Planck} through the SZ effect. The CHEX-MATE census contains two subsamples: the Tier 1, which includes a census of the population at the most recent time (0.05$<z<$0.2); and the Tier 2, the most massive objects to have formed at $z<$0.6 (see Fig. 1 in \citealt{chex-mate21}). Hereafter, we will refer to Tier 1 and Tier2 as  T1 and T2, respectively. Therefore, the CHEX-MATE clusters offers us the opportunity to study the kinematical and dynamical properties of the most massive clusters of the Universe detected through their SZ signal. 

% The mass of galaxy clusters cannot be obtained directly, so scaling relations based on different mass proxies play an important role in this context. 
The masses of galaxy clusters can be obtained directly under specific assumptions \citep[see e.g.][]{pratt19}. Usually, X-ray mass estimates can be recovered either by assuming the hydrostatic equilibrium \citep[see e.g.][]{ettori13} or through scaling relations among integrated quantities, like gas mass and temperature \citep[e.g.][]{Kra06}. The SZ effect can also be used to estimate masses through the Comptonization parameter, Y$_{SZ}$. 

Here, we adopt a dynamical method that uses the line-of-sight velocities of member galaxies and their projected positions with respect to the cluster center to directly measure the mass. Different mass estimation techniques could be biased due to various effects, such as violation of the assumption of  hydrostatical or dynamical equilibrium, or asymmetric, complicated true geometries. To account for these deviations for all these methods, we will study the mass bias parameter $(1-B_1) = M^{SZ}_{500}/M^{M}_{500}$ (\citealt{Ruel2014}; \citealt{Penna17}; \citealt{Amodeo17}), in order to characterise the cluster mass estimations when these are performed using the SZ-X-ray signature or other dynamical techniques. In particular, we will refer to the SZ-X-ray-calibrated mass estimates as $M^{SZ}$, while the mass derived by the \textsc{MG-MAMPOSSt} analysis are identified by $M^{M}$

Our main aims can be summarized in three points: 1) obtain the full kinematic mass profiles for the CHEX-MATE clusters, which allows future comparisons with those obtained from X-ray, SZ and weak-lensing techniques; 2) study the scaling relations and presence of possible biases in the mass estimates derived from different techniques; 3) study statistically the main kinematical properties of the cluster members, by disentangling their velocity anisotropy and orbits depending of the mass of the clusters.

Throughout this paper, we adopt a $\Lambda$CDM cosmology with $\Omega_m=0.3$, $\Omega_{\Lambda}=0.7$, and H$_0=$70 km s$^{-1}$ Mpc$^{-1}$. We further define $r_{\Delta}\equiv r_{\Delta c}$ as the radius of a spherical overdensity within which the mean density is $\Delta$ times the critical density of the universe. The corresponding enclosed mass is $M_{\Delta}$ (\emph{e.g.}, $r_{200},\,M_{200}$ for $\Delta = 200$, and so on). Logarithms in base 10 are defined as: $\text{Log}\, X = \text{log}_{10}\,X$, for any variable $X$, while natural logarithms are indicated by $\ln X$.

\section{Dataset}
\label{sec:data}

\begin{figure}
\centering
\includegraphics[width=\columnwidth]{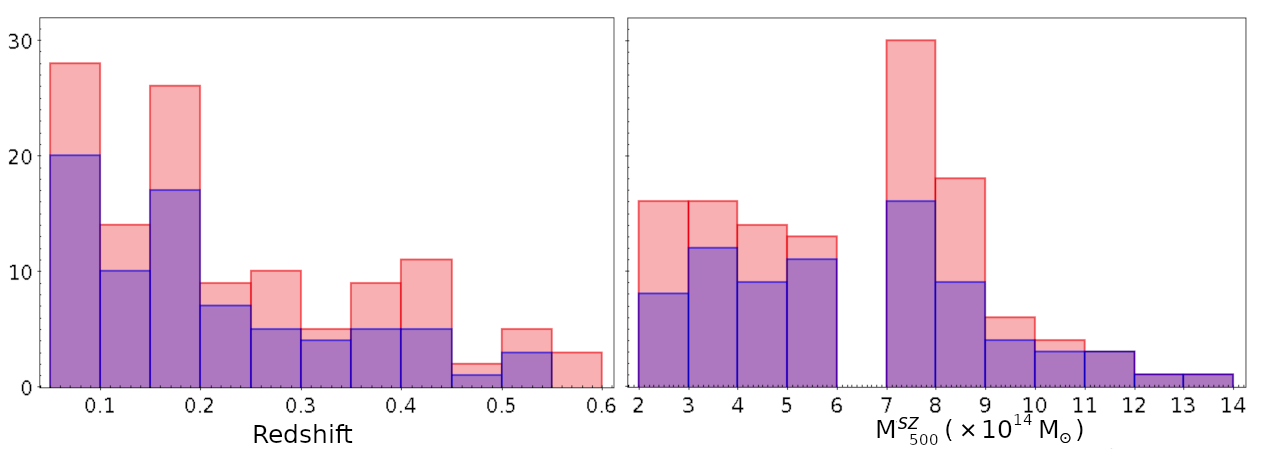}
\caption{Cluster redshift and $M_{500}^{SZ}$ mass distributions of the full CHEX-MATE sample (red) and the considered subsample (blue). $M_{500}^{SZ}$ has been derived iteratively from the $Y_{SZ}-M^{SZ}_{500}$ relation calibrated using hydrostatic masses obtained from XMM-Newton data \citep[see][]{chex-mate21}.}
\label{fig:histo_z_mass}
\end{figure}

Our analysis is based on the radial velocities of cluster members. However, not all the 118 clusters of the CHEX-MATE sample have spectroscopic data of sufficient quality to enable the analysis. So, we restrict our study to 75 clusters, which correspond to a subset of galaxy systems providing at least 50 confirmed members within an aperture radius $R = r_\text{ap,200}$ - as estimated by \cite{Sereno24} - with spectroscopic redshifts, which is a reasonable minimum number of members to obtain an accurate enough analysis of the mass and dynamical properties (see e.g., \citealt{Ferra20}). As stated in \cite{Sereno24}, the cluster centre has been identified as the X-ray luminosity peak. As shown in Fig.~\ref{fig:histo_z_mass}, this subsample represents the 66\% of the full CHEX-MATE clusters, with no bias in redshift or mass. In this Figure the sample is divided into redshift bins (with width equal to 0.05) and mass bins (with width equal to $10^{14}$ M$_\odot$), our analysis covers more than 50\% of the CHEX-MATE clusters in each bin, with the exception of those with $z>0.55$. In other words, the full CHEX-MATE sample is well represented in the present work. Most of the clusters present a large number of members with spectroscopic redshifts, with more than 100 confirmed members within 4.0 Mpc in the majority of the cases, and even more than 300 for some of them. PSZ2 G006.49+50.56 (Abell 2029) and PSZ2 G044.20+48.66 (Abell 2142) represent outstanding cases with more than 1000 total spectroscopic redshifts available. 

The cluster member selection for the sample used in this work was developed by \cite{Sereno24} using the CLEAN method (fully described in \citealt{mamon01}, see Appendix B), which is based on galaxy location in the projected phase-space. So, the radial velocities of cluster members are the same as used in \cite{Sereno24}, who base their analysis on the SDSS-DR18, DESI, and NED databases, in addition to some private catalogues and proprietary data. Moreover, we add new spectroscopic observations\footnote{These new dataset will be made publicly available soon, but it is already available upon request. An extract can be found in Appendix~\ref{app:data}.} for three CHEX-MATE clusters. These are PSZ2 G113.91-37.01, PSZ2G046.10+27.18, and PSZ2G083.29-31.03, all of them in the redshift range $0.37<z<0.41$. We observed galaxies in these fields at the 10.4m GTC telescope with the OSIRIS spectrograph in multi-object observing mode acquiring 2400 s exposures and two mask configurations for each cluster. For more details on the instrumental set-up, data reduction process, and redshift estimates, see \cite{Ferra20} and \cite{Aguado_Barahona_2022}. The procedures applied in this work are exactly the same as those used there. So, we obtain radial velocities for 80, 61 and 66 cluster members, respectively, with a precision of about 70 km s$^{-1}$. These spectroscopic observations are quite limited, showing magnitude limits of $r'\sim 21.5$, and low sample completeness, below 30\% within $r_{500}$. 

For these three clusters we select members by applying a 2.7-$\sigma$ clipping in the phase-space, to minimise the fraction of interlopers\footnote{Galaxies not belonging to the clusters falling inside the projected phase-space.} \citep{Mamon10}. However, we found that only PSZ2G046.10+27.18 has more than $N = 50$ members within the estimated $r_{200}$. From now on, we refer to $r_\text{200,ap}$ as the radius provided by the CLEAN procedure, to distinguish it from the values obtained by the \textsc{MG-MAMPOSSt} fit.

In summary, including the GTC observations, we manage 77 catalogs of cluster member galaxies (one for each cluster), with a total of 17521 radial velocities, which yields a mean of about 220 members per cluster. Among these, 75 clusters have found to provide $N\ge 50 $ confirmed members within the aperture radius $R= r_\text{200,ap}$, after the application of the CLEAN algorithm  (\citealt{Sereno24}). Considering this spectroscopic dataset, we present in the following sections our procedure to analyse the main kinematical and dynamical properties of the CHEX-MATE cluster sample. 

\section{Kinematics analysis setup}
\label{sec:setup}
We infer the cluster mass profiles of the CHEX-MATE sample by employing the \textsc{MG-MAMPOSSt} (Modelling Anisotropy and Mass Profile of Spherical Observed Systems) method (\citealt{Pizzuti:2022ynt}), which is based on the \textsc{MAMPOSSt} \citep{mamon01} algorithm to simultaneously recover the (total) gravitational potential and the orbit anisotropy profile of clusters from kinematic analyses of their member galaxies.

\textsc{MAMPOSSt}  
is based on the assumption that the system is spherically symmetric and that galaxies are collisionless tracers of the gravitational potential. In this case, the dynamics is determined by the so-called (spherical) Jeans' equation:

\begin{equation}\label{eq:jeans}
\frac{\text{d} (\nu \sigma_r^2)}{\text{d} r}+2\beta(r)\frac{\nu\sigma^2_r}{r}=-\nu(r)\frac{\text{d} \Phi}{\text{d} r}\,,
\end{equation}
where $r$ indicates the 3D radial distance from the cluster center,  $\nu(r)$ is the number density profile of galaxies, $\sigma^2_r$ is the velocity dispersion along the radial direction, and $\Phi$ is the total gravitational potential. The quantity
$\beta \equiv 1-(\sigma_{\theta}^2+\sigma^2_{\varphi})/2\sigma^2_r$ is the velocity anisotropy, with $\sigma_{\theta}^2,\,\sigma^2_{\varphi}$ the velocity dispersion components along the tangential and azimuthal directions, respectively. In spherical symmetry, we have $\sigma_{\theta}^2=\sigma^2_{\varphi}$.

\textsc{MAMPOSSt} and \textsc{MG-MAMPOSSt} work in the so-called projected phase-space $(R,v_\text{los})$, PPS hereafter, with $R$ the projected radial distance from the cluster centre - defined to be at the peak of the projected X-ray emission \citep{Sereno24} - and $v_\text{los}$ the line-of-sight (LOS) velocity of each galaxy, computed in the rest frame of the cluster. Given as input the parametric expressions of the gravitational potential, the number density profile, and the velocity anisotropy profile, the code implements a maximum likelihood fit to determine the best set of parameters describing the PPS distribution of cluster members.

\textsc{MG-MAMPOSSt} extends the gravitational potential models available in the original \textsc{MAMPOSSt}, further including generalized models of the velocity anisotropy profile, which we explore here. \textsc{MG-MAMPOSSt} also includes popular modified gravity and alternative Dark Energy scenarios (see \citealt{Pizzuti2021} for a description of the physics), viable at the cosmological level, which we do not analyse in this paper.
Moreover, a recent update of the code implemented a module to include data of the velocity dispersion of the brightest cluster galaxy (BCG) in the fit, when available. Inclusion of the BCG velocity dispersion would allow the method to provide tight constraints on the slope of the dark matter profiles in clusters (see \citealt{Biviano:2023oyf}). However, in this work, we will not use the BCG module.

Therefore, we only consider galaxies outside of a minimum projected radius $R_\text{min} = 0.05\,\text{Mpc}$. The central region is excluded to avoid contamination by the BCG which dominates cluster dynamics at very small radii. Furthermore, in its current implementation, \textsc{MG-MAMPOSSt} assumes negligible streaming motions and a Gaussian 3D velocity distribution of the tracers.

We run \textsc{MG-MAMPOSSt} over our CHEX-MATE subsample of 75 clusters %assuming uninformative, flat priors in each parameter and 
including members up to a maximum projected radius $R_\text{max} = 1.1\,r_{200}^{D}$, where $r_{200}^{D}$ is the estimate of the (dynamical) virial radius coming either from the analysis of \cite{Sereno24} ($r_\text{ap,200}$) or directly from the velocity dispersion (see Sect. 5 in \citealt{Barrena24})
for those clusters for which the former is not available. This definition of maximum radius for cluster membership is a common choice to ensure the validity of the Jeans equation, as $r_{200}$  is  close to the virial radius within Eq.~\eqref{eq:jeans} holds.

We sample the parameter space with a MCMC Metropolis-Hastings algorithm for 110~000 points, considering the first 10~000 as the burn-in phase.
Our final chain contains 100~000 samples for each run; this number has been chosen has a reasonable value to save computational time while ensuring convergence. In more detail, we checked convergence by considering the four clusters with the fewest members ($N< 60$) within $r_\text{ap,200}$; and running $n = 5$ test chains for each of them. We then computed the corresponding Gelman-Rubin diagnostic coefficients $\hat{R}$ \cite{Gelman92}\footnote{the Gelman-Rubin coefficient is defined as $\hat{R} = [W(L-1)/L + B/L]/W$, where $L$ is the length of the chains, $B$ the variance of the means of the chains, and $W$ the averaged variances of the individual chains across all chains.}
checking that the requirement $\hat{R} \le 1.1$ is always satisfied, as done in \cite{Pizzuti_2024b}.

As for the main analysis, a total of eight \textsc{MG-MAMPOSSt} chains are generated per cluster in the sample, exploring the combination of two models for the number density distribution and four models for the total mass profile, which we list in the following. The Navarro-Frenk-White (NFW) model can be written as,
%-==================================
\begin{equation}
\label{eq:Mass_NFW}
    M_\text{NFW}(r) = M_{200}\frac{\ln (1+r/r_\text{s}) -r/r_\text{s}/(1+r/r_\text{s})}{ \ln(1+c_{200}) - c_{200}/(1+c_{200})}\,, 
\end{equation}
%-==================================

where $M_{200} =200\,(H^2(z)/2G)\,r^3_{200}$ is the enclosed mass at $r_{200}$ and $c_{200} = r_{200}/r_\text{-2}$ is the concentration, with $r_\text{-2}$ being the radius at which the logarithmic slope of the density profile is $-2$.  For the NFW model, this radius corresponds to $r_\text{s}$.\\

The Hernquist profile has a steeper decline of the density at large radii, and the mass enclosed in a radius $r$ reads (\citealt{Hernquist01}), 
\begin{equation} \label{eq:mass_Her}
    M_\text{Her}(r) = M_{200}\frac{r_\text{s}^2+ r_{200}^2}{r_{200}^2}\frac{r^2}{r+r_{200}^2}\,,
\end{equation}
where $r_\text{s} = 2\, r_{-2}$.

The Burkert model \citep{Burkert01} has a central core in the density distribution, and the mass is given by 
\begin{equation}
\begin{split}
      &  M_\text{Bur}(r) =  \\
     &\frac{M_{200}\left[\text{ln}  \left((r/r_\text{s})^2+1\right)+2 \text{ln}  (r/r_\text{s}+1)-2 \tan ^{-1}(r/r_\text{s})\right]}{\left[\text{ln}  \left((r_{200}/r_\text{s})^2+1\right)+2 \text{ln} (r_{200}/r_\text{s}+1)-2 \tan ^{-1}(r_{200}/r_\text{s})\right]}\,,\\
    \end{split}
\end{equation}
with $r_\text{s} \simeq r_{-2}/1.5 $.

Finally, the Einasto model \cite{Einasto1965} can be written as,
\begin{equation}
    M_\text{Eis}(r)= M_{200}\frac{\gamma\left[\frac{3}{m}, \frac{2}{m}\left(\frac{r}{r_s}\right)^m\right]}{\gamma\left[\frac{3}{m}, \frac{2}{m}\left(c_{200}\right)^m\right]},
\end{equation}
where we set the shape parameter\footnote{We checked that varying the exponent between $m=2$ and $m=5$ does not lead to statistically relevant changes in the final posterior.} $m = 3$ and $r_\text{s} = r_{-2}$ as for the NFW profile. Therefore, all models are described by two free parameters, the mass $M_{200}$ (or equivalently, the radius $r_{200}$) and the scale radius $r_\text{s}$, which will be fitted within the \textsc{MG-MAMPOSSt} algorithm.

As mentioned above, we further consider two models for the number density profile: the projected NFW and the projected Hernquist (pNFW and pHer, respectively). These models are defined as the integral of Eqs.~\eqref{eq:Mass_NFW},\eqref{eq:mass_Her} along the LOS. While the profiles are characterized by two free parameters, only the scale radius $r_\nu$\footnote{we define the scale radius of the number density profile as $r_\nu$ to avoid confusion with that of the mass profile} is relevant in the analysis, as the other factors out in the solution of the Jeans equation implemented in \textsc{MG-MAMPOSSt}. Note that for $r_\nu$, we first perform a preliminary fit to the member galaxy distribution in the phase space which does not require the binning of data \citep{Sarazin80}, assuming a constant completeness of the sample in the projected radial distance explored. This last assumption is  rather strong, since our dataset exhibits quite incomplete and inhomogeneous spatial distributions; however, we carried out a detailed analysis on a few clusters to check that reasonable variation of the completeness does not provide relevant changes in the reconstructed mass profile.
We then use the 95\% C.L. limits of the inferred $r_\nu$ as the upper and lower bounds for the flat prior range in the \textsc{MG-MAMPOSSt} fit (see below). We have further checked that varying this interval produces negligible effects in the reconstructed mass profile.

As for the velocity anisotropy profile, we adopt a generalized Tiret (gT hereafter, see \citealt{Mamon19}) model, which is an extension of the Tiret profile \citep{Tiret01} able to capture a quite broad ranges of possibilities for the orbits of member galaxies in clusters:
\begin{equation}
        \beta_{gT}(r) = \beta_0 +\beta_{\infty} r/(r+r_{\beta}), \label{e:tiret}
\end{equation}
where $\beta_0$ and $\beta_\infty$ are the inner and outer values of the anisotropy, respectively, and $r_{\beta}$ is a characteristic scale radius.
In all cases, we set $r_{\beta}=r_\text{-2}$ (i.e., the scale radius of the mass profile) as suggested by numerical simulations \citep{Mamon10}.
In \textsc{MG-MAMPOSSt}, we work with the rescaled parameters $\mathcal{A}_{0 ,\infty} = (1 - \beta_{0 , \infty})^{-1/2}$, which are equal to one for completely isotropic orbits.

We adopt flat priors on all the fitted parameters, $r_{200}/\text{Mpc} \in [0.5,5.0]$, $r_\text{s}/\text{Mpc} \in [0.04,4.0]$, $\mathcal{A}_{0,\infty} \in [0.4,7.0]$, $r_\nu \in [r_\nu^\text{low},r_\nu^\text{up}]$, where the last range is between the 95\% lower and upper limits from the external fit of the projected number density distribution. Except for $r_\nu$, the priors are non-informative.

For each posterior, we apply the Bayesian Information Criterion (BIC hereafter, see, e.g., \citealt{Mamon19}) to quickly select the best model among the eight combinations analysed:
\begin{equation}
    \text{BIC} = k\ln\,N - 2\ln \mathcal{L}\,,
\end{equation}
where $k$ is the number of free parameters, $N$ the number of data-points and $\mathcal{L}$ the likelihood at the best fit. Generally, a model is said to be strongly preferred when its BIC is lower by a factor of $\sim 6$ with respect to the others.  For the models with the same number of free parameters, the BIC comparison is equivalent to the chi-square $\chi^2$ comparison. It is important to stress that \text{MG-MAMPOSSt} only indicates which model is favored relative to the others, but not the best model in an absolute sense.
We further quantify the systematics induced by the modeling as half of the maximum variation between all the best-fit values of each parameter. 

\section{Results}
Here we present the main outcomes from the \textsc{MG-MAMPOSSt} analysis for the CHEX-MATE sample of 75 clusters. 
Among the mass ansatz, all the models adequately fit the PPS distribution, with the NFW profile performing better for 45\% of the clusters, followed by Burkert (32.5\%), Einasto (12.5\%), and Hernquist (8\%). This distribution may indicate a slight preference towards steeper central profiles with respect to those with cores.  While it is true that the presence of interlopers that survived the CLEAN procedure can produce a steeper profile, thus biasing in favor
of NFW-like models, the statistical significance of the latter is in general negligible, with very small $\Delta$BIC ($\lesssim 1.0$ on average). In contrast, the fits to the number density distribution of galaxies notably prefer the projected Hernquist model compared to the NFW model for more than 70\% of the analysed clusters, with several cases where $\Delta$BIC$ \gtrsim 6.0$. This preference seems to be in contrast to other studies (e.g. \citealt{van_der_Burg_2016}), which indicate a preference of the pNFW model. However, one should note here that the assumption of a constant completeness may play a significant role. Indeed, while it is true that the constraints on a mass profile distribution are not strongly affected by the choice of the model of the number density profile in the \textsc{MG-MAMPOSSt} fit, the slope of $\nu(r)$ and the value or the scale radius $r_\nu$ depends on the completeness.  In order to deal with this issue, one should compute the completeness, cluster by cluster, by comparing photometric and spectroscopic catalogs. However, this analysis is beyond
the scope of the paper. As an example, we consider the case of the cluster PSZ2 G049.22+30.8, for which the completeness as a function of the projected radius has been estimated according to the model
\begin{equation}
     C(R) =A\exp(B\,R)\,,
\end{equation}
with $A = 0.817$, $B= -0.413$. While in this case the best fit projected number density is always a NFW profile, the value of the scale radius changes from $r_\nu = 0.31^{+0.18}_{-0.10}$ when accounting for the correct completeness, to  $r_\nu = 0.59^{+0.10}_{-0.13}$, when a constant completeness is assumed (uncertainties are given at 95\% C.L.).

 In Table \ref{tab:results} we report the regression results.
In the  second and third columns, we list the best fit models (i.e., the models that give the highest posterior) for mass and density, respectively. For each entry, the first uncertainties refer to the (statistical) 95\% confidence intervals, while the second values are the systematic uncertainties computed as discussed above. The last four columns show the value of the log posterior $\ln P$ of the best fit, the cluster redshift $z$, the value of the Anderson-Darling coefficient $A^2$ (see Sect. \ref{sec:structure}), and the estimated line-of-sight velocity dispersion at $r_{200}$, $\sigma_\text{ap,200}$. Note that the \textsc{MG-MAMPOSSt} procedure is pretty robust in estimating $r_{200}$ against the choice of the mass, anisotropy and number density models, as shown by the small systematic uncertainties compared with the statistical errors.

\begin{figure}
\centering
\includegraphics[width=0.7\columnwidth]{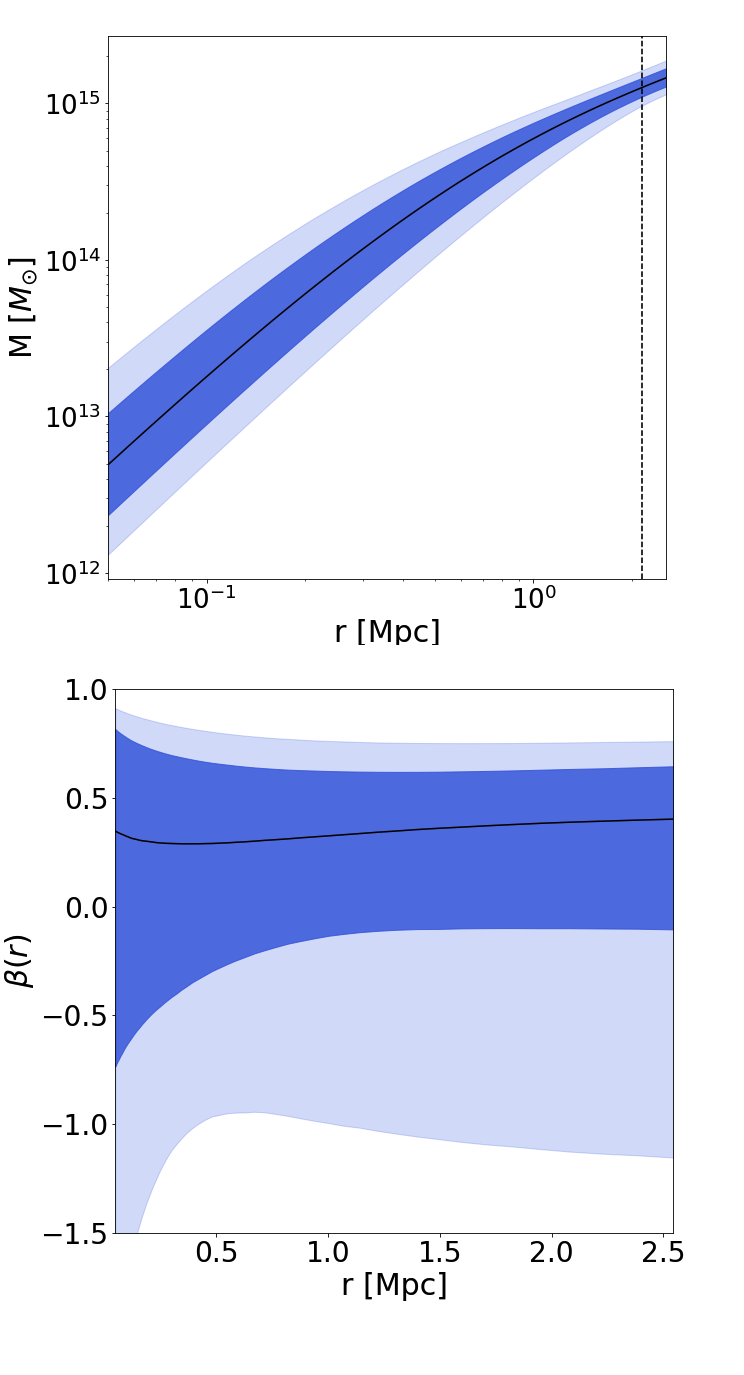}
      \caption{Mass profile (top) and radial velocity anisotropy profile (bottom) of the cluster PSZ2~G067.17+67.46, as obtained by the \textsc{MG-MAMPOSSt} analysis with $N_\text{gal} = 299$ galaxies adopting  a NFW and a gT models, respectively. In both plots, darker and lighter shaded regions refer to 68\% and 95\% confidence levels, respectively. The vertical dashed line in the top panel identifies the best fit for $r_{200}=2.19\,\text{Mpc}$.}
\label{fig:exampleAnis}
\end{figure}

In Fig. \ref{fig:exampleAnis} we show an example of the mass and anisotropy profiles obtained from \textsc{MG-MAMPOSSt} for one rich cluster ($N_\text{gal} = 291$ spectroscopic members within $r_\text{ap,200}$) in the sample, PSZ2~G067.17+67.46. The mass is parametrized as an NFW model, which gives the best-fit profile, with a pNFW model for the number density of the member galaxies.

\begin{figure}
\includegraphics[width=0.8\columnwidth]{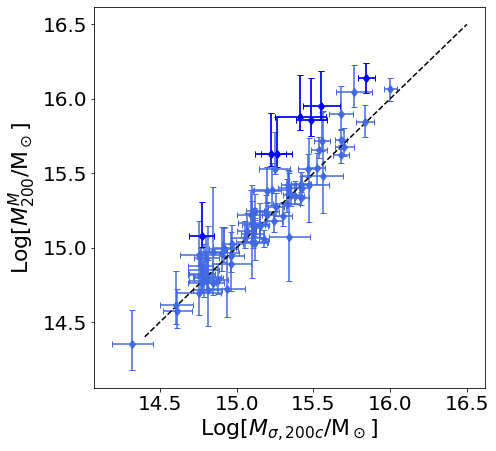}
      \caption{Scatter plot of  $\text{Log}\, M_{200}$, inferred from  the kinematic analysis with \textsc{MG-MAMPOSSt} (vertical axis), and $\text{Log}\,M_{\sigma,200c}$ obtained by \cite{Sereno24} (horizontal axis). The errorbars represent the 68\% C.L, while the black dashed line indicates the bisector. Darker points highlight those clusters fro which $\text{Log}\,(M_{200}^M/M_{\sigma,200c}) > 0.3$.  }
\label{fig:MassM_vsMassS}
\end{figure}

We then compute the mass inside the virial radius as  $M_{200} = (100\,H^2(z)/G)\times r_{200}^3$, and we perform a first comparison with the results of \cite{Sereno24}. Fig. \ref{fig:MassM_vsMassS} displays the values of $\text{Log}\, M_{200}$ as obtained by \textsc{MG-MAMPOSSt} (vertical axis), and $\text{Log}\,M_{\sigma,200c}$ estimated from the aperture velocity dispersion $\sigma_\text{ap,200}$ with the method of \cite{Sereno24} (horizontal axis). The superscript $(M)$ refers to the \textsc{MG-MAMPOSSt} estimates.

\begin{figure}
\includegraphics[width=0.8\columnwidth]{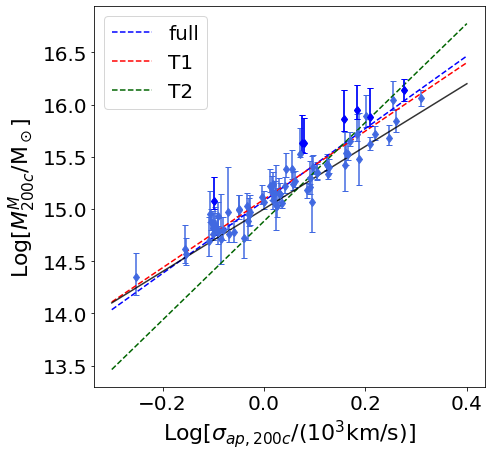}
      \caption{Logarithm of the aperture velocity dispersion within $R= r_{200}$, $\sigma_\text{ap,200}$ compared to the logarithm of $M_{200}^M$ estimated from \textsc{MG-MAMPOSSt}. The bars refer to 68\% uncertainties. The blue line indicates the best fit linear model of eq. \eqref{eq:fitmsigma} when $z = 0.2$. The black solid line identifies the $\sigma \propto M^{3}$ relation.  
              }
         \label{fig:Msigma}
\end{figure}

The results show an overall good agreement among the two estimates, confirming the robustness of the simpler analytical approach of \cite{Sereno24}. However, for seven clusters \textsc{MG-MAMPOSSt} tends to overestimate $M_{200}$ with respect to the findings of \cite{Sereno24}. For those clusters we found an average logarithmic difference $\text{Log}\, M_{200}^M - \text{Log}\,M_{\sigma,200c} > 0.3 $ dex.%One of them, PSZ2G056.77+36.32, exhibits clearly disturbed features, as we will discuss more in detail below (see also the plots in the first row of Fig.\ref{fig:badpps}).

If we look at the relation between $M_{200}^{M}$ and the aperture velocity dispersion $\sigma_\text{ap,200}$ (Fig.~\ref{fig:Msigma}) obtained by \cite{Sereno24} using the CLEAN method (\citealt{mamon01}), it is easy to see that six of the seven clusters are characterized by large values of $\sigma_\text{ap,200} > 10^3 $ km/s. 
This pattern suggests that clusters in the CHEX-MATE sample with very large velocity dispersion may suffer a stronger contamination of interlopers. This fact was first found by \cite{Mamon10} who, working with simulated halos, found that the fraction of interlopers should be (slightly) more important in the more massive halos since they have (slightly) lower concentrations (\citealt{NFW97}; \citealt{Maccio08}) and they typically live in richer environments. That is, the larger fraction of interlopers that survived the membership selection performed by \cite{Sereno24} in high-mass clusters could translate into a higher mass estimation by \textsc{MG-MAMPOSSt} - which is a more complex model with respect to that of \cite{Sereno24} - for the most massive systems. 

Moreover, in \textsc{MG-MAMPOSSt} the concentration $c_{200} = r_{200}/r_\text{-2}$ is a free parameter (or better, it is a quantity derived from two parameters optimized within the fit), while in \cite{Sereno24} it is fixed as a function of $r_{200}$ using the concentration-mass relation proposed by \citet{Diemer19}.

As shown by the blue points in Fig. \ref{fig:Cm}, the off-trend clusters are characterized by a concentration estimated by \textsc{MG-MAMPOSSt} that is lower than the prediction of \cite{Diemer19} (brown solid line), consequently boosting the reconstructed $M_{200}$.

%The existence of a tight relation between mass and velocity dispersion for clusters is well known in literature (\citet{Munari2013,Saro13}). Here we aim at investigating the relation for the CHEX-MATE sample, using the values of $M_{200}$ given by the best fit of the \textsc{MG-MAMPOSSt} analysis and the aperture velocity dispersions $\sigma_\text{ap,200}$ provided by the analysis of \cite{Sereno24}. The relation is shown in the top panel of Figure \ref{fig:Msigma}.
The blue line in Fig. \ref{fig:Msigma} shows the best orthogonal distance regression (ODR) fit for the linear model
\begin{equation}
\label{eq:fitmsigma}
\text{Log}\left(F_z\frac{M_{200}^M}{10^{14}\,\text{M}_\odot}\right) = \alpha_\sigma\, \text{Log}\left(\frac{\sigma_\text{ap,200}}{10^{3}\text{km}\,\text{s}^{-1}}\right) + \beta_\sigma + \gamma _\sigma\,\text{Log} F_z\,,
\end{equation}
when fixing the redshift to the pivot value of the sample $z_p = 0.2$; in eq. \eqref{eq:fitmsigma}, $F_z = H(z)/H(z_p)$. We found: 
\begin{equation}
    \alpha_\sigma = 3.61 \pm 0.17,\,\,\,\,\,\,\, \beta_\sigma = 1.07 \pm 0.01, \,\,\,\,\,\,\,
    \gamma_\sigma =-0.83 \pm 0.70\,,
\end{equation}
which is slightly steeper than $\alpha \sim 3$, as in e.g., \cite{Munari2013,Ferra20}. If we consider the T1 and T2 clusters separately, the ODR fit of Eq. \eqref{eq:fitmsigma} provides a slightly lower $\alpha^{T1} = 3.31\pm0.17$ for the Tier 1 sample - close to the $\sim \sigma^3$ relation and a higher value $\alpha^{T2} = 4.81\pm0.64$ for Tier 2. This reflects the fact that T2 contains massive clusters by construction, and that among the seven off-trend objects, five belong to T2.
The best ODR fits for $z_p$ are shown as the red and green dashed lines for T1 and T2 respectively.

\subsection{Comparison with SZ mass estimates}

In order to compare our kinematic masses with the ones obtained from the SZ effect, we further derive the distribution of $M_{500} = 500\,H^2(z)/(2G)\times r_{500}^3 $, as follows: for each value of $r_{200},r_\text{s}$ in the MCMC chain we obtain the corresponding parametric mass profile $M(r|r_{200},r_\text{s})$. The radius $r_{500}$ is then given by the numerical solution of the equation
\begin{equation}
    \frac{M(x)}{x^3} = \frac{500\,H^2(z)}{2G}\,.
\end{equation}
From that, the computation of $M_{500}$ is straightforward.

\begin{figure}
   \centering
   \includegraphics[width=0.8\columnwidth]{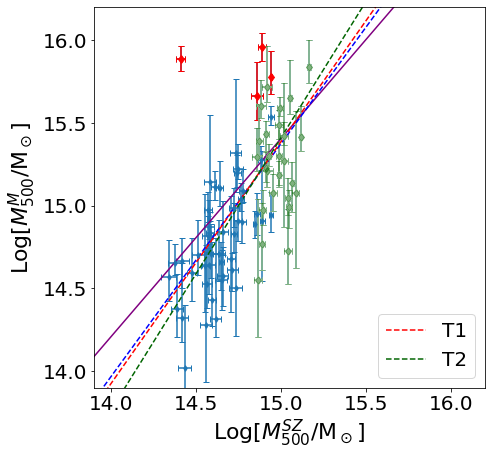}
      \caption{Scatter plot of $M_{500}$ inferred from SZ-effect by Planck (on the vertical axis) and from the kinematic analysis with \textsc{MG-MAMPOSSt} (horizontal axis). The errorbars represent the 68\% C.L. while the red points show those clusters for which the logarithmic difference ($\delta \eta$) is larger than 0.8 dex. Dark-green points highlight clusters belonging to Tier 2. The purple line refer to a ODR fit considering all the clusters, while the blue one is the linear fit done by discarding the four above mentioned clusters. The best fits (with the four outliers removed) linear models for T1 and T2 separately are further shown as the red and dark-green dashed lines, respectively.
              }
         \label{fig:MszMm}
   \end{figure}

In Figure \ref{fig:MszMm} we plot the values of $\text{Log}\, M_{500}$ as obtained by \textsc{MG-MAMPOSSt} (vertical axis) and inferred by SZ  (horizontal axis), where the superscript $(M)$ and $(SZ)$ refers to the \textsc{MG-MAMPOSSt} and SZ estimates respectively. We found a quite strong correlation between the two estimates (Pearson coefficient $r = 0.68$, p-value $< 10^{-5}$); we fit the linear relation
\begin{equation}\label{eq:logmfit}
    \text{Log} \left(\frac{M^{SZ}_{500}}{10^{14}\,\text{M}_\odot} \right)= \alpha_{SZ}\,\text{Log} \left(\frac{M^{M}_{500}}{10^{14}\,\text{M}_\odot}\right) + \text{Log}(1-B)\,,
\end{equation}
where the mass-dependent bias (1-B) is defined similarly to that done, for example, in \cite{Ferra21} and \cite{Aguado_Barahona_2022}. This way, the usual bias $(1-B_1) = M^{SZ}_{500}/M^{M}_{500}$ is recovered when $\alpha_{SZ} = 1$.

We obtain $\alpha_{SZ} = 0.84\pm0.05$, while an overall mass-dependent bias $\text{Log}(1-B) = -0.18\pm 0.06$ dex is found when all the galaxy clusters in the sample are considered, corresponding to $(1-B) = 0.67 \pm 0.09$; this means that the kinematic results from \textsc{MG-MAMPOSSt} produce larger masses with respect to those derived from the  \textit{Planck} SZ signal. This value seems to be in agreement within errors with those obtained by \citet{Planck_res_vi} and \citet{Lesci23} from BAO and 3D clustering of the PSZ2 clusters. In addition, our estimate is similar to that found by the fit of \cite{Sereno24}, although slightly larger than that of \cite{Ferra21} and \cite{Aguado_Barahona_2022}, who obtained a bias of $(1-B) \sim 0.80$ comparing Planck SZ-X-ray calibrated masses of 270 PSZ1 and 388 PSZ2 clusters, respectively, with the kinematic mass estimated through the total velocity dispersion. 
When forcing $\alpha_{SZ}$ to unity, however, the difference among the two mass estimates is even larger, almost doubled: $(1-B_1) = 0.43\pm 0.08$, which is not consistent with the above mentioned works.

% This suggests that mass reconstructions using kinematics of member galaxies tend in general to % moderately overestimate the mass with respect to analyses involving the Intra-Cluster Medium. 
%----------------------------------------------------------- Mass comparison
%
%______________________________________________________________

It is interesting to note that the sample includes a few clusters for which the kinematic mass is about one order of magnitude larger than the one given by the SZ proxy. We marked four clusters for which 
\begin{equation}
    \delta \eta = \text{Log}\, M^{M}_{500} -\text{Log} \,M^{SZ}_{500} > 0.8\,,
\end{equation}
as red points in Figure \ref{fig:MszMm}. However, the PPS distributions of those clusters show evident sign of a disturbed dynamical state, as displayed in the left plots of Figure \ref{fig:badpps}, with very large velocity dispersions and presence of possible important structures that are not part of the main cluster, as in the case of PSZ2G056.77+36.32, and PSZ2G040.03+74.95. 

\begin{figure}
\centering
\includegraphics[width=0.7\columnwidth]{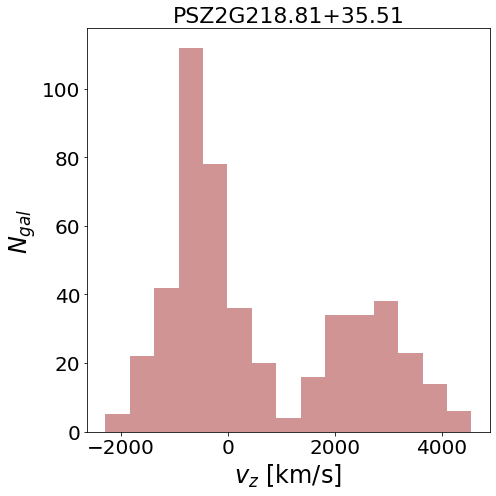}
      \caption{LOS velocity distribution for the system PSZ2G218.81+35.51, made of two separate clusters, as mentioned in  \cite{Sereno24}. 
              }
         \label{fig:109}
\end{figure}

The disturbed dynamical state of these four clusters is further indicated by the distributions of the velocities along the line of sight (central plots of Figure \ref{fig:badpps}), that for PSZ2G056.77+36.32 and PSZ2G040.03+74.95 exhibit a noticeable bimodal distributions. Note that, here we are considering the members selected by running the CLEAN algorithm in its default settings. However, as already done in  \cite{Sereno24}, a refined analysis can be performed to study these systems. In more detail, PSZ2G040.03+74.95 - aka  Abell 1831 - has been already identified as a double system: it is made of two clusters A,B at $z = 0.063, 0.076$ respectively, nearly aligned along the line-of-sight, as discussed in \cite{Sereno24}. A similar situation holds for the object PSZ2G218.81+35.51 (see \cite{Sereno24} and also the histogram of the LOS velocities in Figure \ref{fig:109}). We found a logarithmic difference $\delta \eta = 0.68 \pm 0.42$ for PSZ2G218.81+35.51, where the large uncertainties reflect the inefficiency of \textsc{MG-MAMPOSSt} in fitting the double system with a single mass model. 

By considering the two sub-groups in PSZ2G040.03+74.95 and PSZ2G218.81+35.51 as independent clusters\footnote{The selection of members for the subsystems is performed as discussed in \cite{Sereno24}}, the kinematical analysis gives $M_{200} < 7 \times 10^{14} \, M_\odot$ and $M_{200} = 53^{+9 }_{-26}\times 10^{14} \, M_\odot$ for PSZ2G040.03+74.95 A, B, and $M_{200} = 2.4^{+1.8 }_{-2.2} \, M_\odot$, $M_{200} = 6.1^{+3.6 }_{-2.7} \, M_\odot$ for PSZ2G218.81+35.51 A, B. Note that for PSZ2G040.03+74.95 A we are able only to set an upper limit with \textsc{MG-MAMPOSSt}, given the small number of members considered in the fit  (only 25 members lie within $R=r_{200}$, well below the limit of 50 we imposed on the full sample) and the relatively small $\sigma_\text{ap,200} = 461$ km s$^{-1}$, compared to the error in the line-of-sight velocities.
It is important to mention that the results for these bi-modal systems are in agreement (despite with larger uncertainties) with the values obtained by the simple halo model of \cite{Sereno24}.\\
For the most massive sub-components in the bimodal clusters (which are identified as the CHEX-MATE clusters, see \cite{Sereno24} and references therein), the  logarithmic difference is now reduced to $\delta\eta = 0.8 \pm 0.3$ (PSZ2G040.03+74.95) and $\delta\eta = 0.15 \pm 0.11$ (PSZ2G218.81+35.51), instead of 1.4 and 0.68. 
These interesting cases, as well the other three clusters mentioned above, deserve a separate detailed analysis of their properties. Such an analysis is beyond the scope of this paper and it will be performed in a separate work. From now on,  we will consider only the main sub-components of the double systems in the forthcoming analyses.\\
 If we remove the four clusters with large $\delta \eta$ and we fit again eq. \eqref{eq:logmfit} (blue dashed line in Figure \ref{fig:MszMm}), we remarkably find a negligible mass-dependent bias within the uncertainties $\text{Log} (1-B) = 0.02 \pm 0.04$, which corresponds to $1-B = 1.04 \pm 0.09$, with a slightly flatter slope $\alpha_{SZ} = 0.71 \pm 0.04$. Note that, however, the bias is still quite large when fixing the slope to $1$,  $1-B_1 = 0.54 \pm 0.11$, from which we can conclude that kinematic mass estimates are a factor of two larger, on average, with respect to SZ probes. Note that this value is now in agreement with the findings of, e.g., \cite{Lesci23}. Here we have ignored the possible redshift dependence of the sample and neglected the Malmquist bias, both were already considered in \cite{Sereno24}. In particular, SZ masses are biased high, which may produce an underestimation of $B$. When considering separately T1 and T2 (after removing the four biased clusters) the fit is still consistent with a negligible mass-dependent bias; in particular,  $(1-B) = 1.12 \pm 0.13$ for the former, while the latter provides $(1-B) = 1.32 \pm 0.44$. The larger uncertainties in Tier 2 reflect a quite wide scatter in $\delta\eta$, as shown by the green points in Fig.~\ref{fig:MszMm}. For $B_1$, we found $(1-B_1) = 0.64 \pm 0.04$ for Tier 1 and $(1-B_1) = 0.40 \pm 0.04$ for Tier 2.

\begin{figure}
\includegraphics[width=0.9\columnwidth]{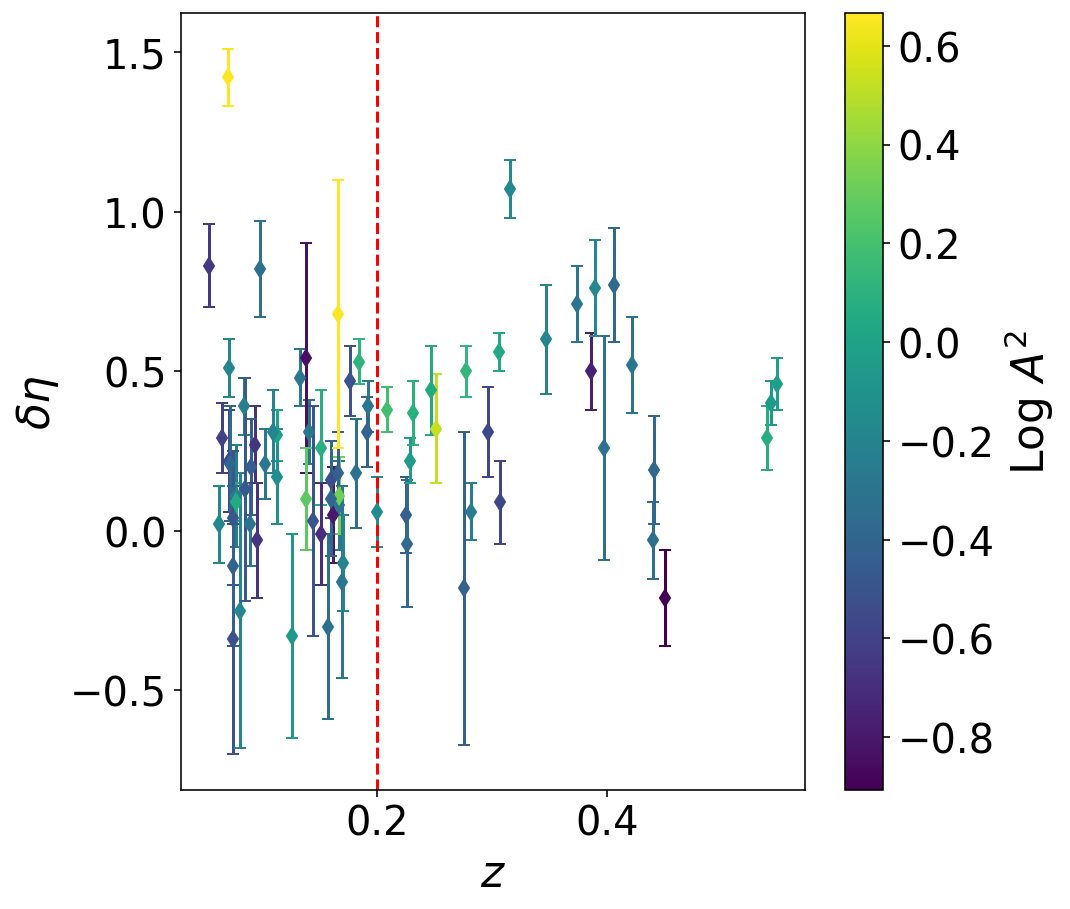}
      \caption{$\delta \eta \equiv \text{Log} (M_{500}^M/M_{500}^{SZ})$ as a function of the redshift. The error bars indicate the 68\% statistical uncertainties. Points are color-coded according to the values of the $A^2$ coefficient. The red vertical line indicates the pivot redshift $z=0.2$}
         \label{fig:z}
\end{figure}

Note that $\delta \eta$ does not exhibit a strong correlation with redshift (see Fig. \ref{fig:z}), with a quite scattered distribution. However, if we compute the fraction of galaxies with $\delta \eta > 0.5 $ in two redshift bins, $z < 0.2 $ and $z> 0.2$, we find 0.14 and 0.28, respectively. This difference should not come as a surprise, since clusters with $z>0.2$ belong to the the Tier 2 selection, which contains massive clusters by construction (thus characterized by larger velocity dispersions, see, e.g., \citealt{Sereno24}). Those clusters exhibit overall larger discrepancies between $M^{SZ}_{500}$ and $M^{M}_{500}$, attributed to a potentially stronger contribution of interlopers in the kinematics analysis, as already mentioned above, in clusters with high velocity dispersions. 
%We highlight in green those clusters for which $\sigma_\text{ap,200} > 1200$ km s$^{-1}$ in Fig. \ref{fig:z}.

\subsection{Concentration-mass relation}
\begin{figure}
\includegraphics[width=0.9\columnwidth]{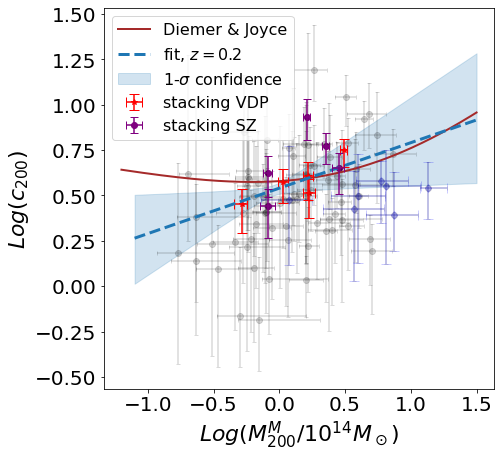}
      \caption{Concentration-mass relation for the CHEXMATE sample. The errorbars indicates 68\% uncertainties in the parameters; we highlidhted in light blue the clusters with $\text{Log}(M^{M}_{200}/M_{\sigma,200c})> 0.3$. The blue-dashed line is the best fit Bayesian linear regression derived with the method of \cite{se+et15_comalit_I}; the shaded region represents the corresponding $1\,\sigma$ contour.
      Purple and red crosses indicates the values obtained by stacking clusters in bin of SZ mass and $\sigma_\text{ap,200}$, as explained in Sect. \ref{sec:anis}. The brown solid line shows the theoretical prediction of \cite{Diemer19}.
              }
         \label{fig:Cm}
\end{figure}

In Fig.~\ref{fig:Cm} we plot the relation between $c_{200}$ and $M_{200}$ of the sample, as derived by the \textsc{MG-MAMPOSSt} analysis. For the two double systems PSZ2G056.77+36.32 and PSZ2G218.81+35.51 we consider here separately the sub-components, further discarding PSZ2G056.77+36.32 A, given its limited statistics. 
Although the points appear quite scattered and have large uncertainties, a mild, positive correlation can be observed (Spearman rank coefficient $0.33$, corresponding to a p-value of $2\times 10^{-3}$).

For a quantitative analysis, we perform a Bayesian regression.  We follow the CoMaLit (Comparing Masses from Literature) scheme described in \citet{se+et15_comalit_I, ser+al15_comalit_II, se+et15_comalit_IV, se+et17_comalit_V}, which we refer to for details, and implemented in the \textsc{R}-package \texttt{LIRA} \citep{ser16_lira}.\footnote{The package \texttt{LIRA} (LInear Regression in Astronomy) is publicly available from the Comprehensive R Archive Network at \url{https://cran.r-project.org/web/packages/lira/index.html}.}

The mass - concentration relation is modeled with a linear relation in log-space, similarly to eq.~\eqref{eq:fitmsigma},
\begin{equation}
\label{eq_scaling_1}
\text{Log} (c_\text{200}) = \alpha_C + \beta_C \,\text{Log}\, (M^\text{M}_\text{200} / M_\text{pivot}) + \gamma_C\, \text{Log} \,F_z    \, ,
\end{equation}
where $F_z = H(z) / H (z_\text{ref})$ is the redshift dependent Hubble parameter normalised to the reference redshift of the sample, $z_\text{ref} = 0.2$, and $M_\text{pivot} = 10^{15}M_\odot$ is the pivot mass. The concentration at the reference mass and redshift is given by $c_\text{200,ref} =10^{\alpha_C}$.

Statistical biases and correlation effects have to be carefully dealt with \citep{ser+al15_cM}. For our analysis, we assume that both $X = \text{Log} (M^\text{M}_\text{200c} / M_\text{pivot})$, that is, the covariate in Eq.~(\ref{eq_scaling_1}), and $ Y = \text{Log}(c_\text{200})$, that is, the response, are intrinsically scattered with respect to the true mass, $Z$, whose distribution we model as a Gaussian with redshift dependent mean and variance \citep{se+et15_comalit_IV}. We assume that the covariate is scattered but unbiased with respect to the true mass. We consider heteroscedastic and correlated uncertainties for mass and concentration. For central location and scale, we consider the bi-weight estimators  \citep{bee+al90}.

We apply non-informative priors \citep{se+et15_comalit_IV, se+et17_comalit_V}: the uniform distribution for $\alpha_C$ and the mean of the distribution of $Z$; the Student's $t_1$ distribution with one degree of freedom for the slopes $\beta_C$ and $\gamma_C$; the Gamma distribution for the inverse of either variances or intrinsic scatters.

The regression favors a flattening and upturn of the relation at large masses, in agreement with theoretical predictions \citep{Diemer19}. In particular, the best fit values of the linear relation are 
\begin{equation}
\alpha_C = 0.49 \pm 0.07,\, \,\,\,\,\,\, \beta_C = 0.20 \pm 0.31, \,\,\,\,\,\,\, \gamma_C = - 0.23 \pm 1.6.   
\end{equation}

To test the robustness of the results on the regression scheme, we also fit the relation with a simple linear model, similarly to that done, for instance in \cite{Biviano17}, using the ODR method and neglecting the redshift dependence, 

\begin{equation}
    \text{Log}\,c_{200} = A_1\, \text{Log}\,\left(\frac{M^{M}_{200}}{10^{14} M_\odot}\right) + A_2\,.
\end{equation}

We found $A_1 = 0.27 \pm 0.08$, $A_2 = 0.20 \pm 0.11$, which confirms  the trend of rising concentration for high mass clusters. The same is valid if we fit separately T1 and T2 clusters, in particular $A^{T1}_1 = 0.35 \pm 0.12$ and $A^{T2}_1 = 0.21 \pm 0.14$.  Our findings might appear to be in tension with other theoretical expectations (e.g., \citealt{Dutton14,Ragagnin_2020}) and observational determinations of the $c-M$ relation (e.g., \citealt{Merten15,Biviano17}, where the latter is also based on the \textsc{MG-MAMPOSSt} technique). However, this statement should be considered cautiously. The CHEX-MATE sample covers the very massive end of the halo mass function at intermediate redshifts, where the upturn is visible. Selection effects  as well as the uncertainties in the kinematic analysis may play a significant role; in particular, X-ray and SZ-selected samples tend to favor more centrally concentrated and relaxed clusters, biasing the inferred c-M relation upward (e.g., \citealt{Meneghetti14}). Moreover, as mentioned before, the presence of residual interlopers may affect the high-mass end of the sample, giving rise to the observed trend.

\subsection{Substructures and un-relaxation}
\label{sec:structure}
To further investigate the overall kinematical state of the cluster sample, we compute the Anderson-Darling coefficient $A^2$ (\citealt{Anderson52}), which measures deviation from Gaussianity of the LOS velocity distribution, which is a good indicator of the dynamical state of a cluster (e.g., \citealt{Roberts18}). Large values of $A^2$ have been shown to be associated with deviations from relaxation (e.g., \citealt{Pizzuti19b,Barrena24}). The colors of each point in Fig.~\ref{fig:z} indicates the corresponding value of the (logarithm of the) Anderson-Darling coefficient, computed using all galaxies within $R = r_{\text{ap},200c}$ as given by CLEAN (\citealt{Sereno24}). 
%\begin{figure}
%\includegraphics[width=0.9\columnwidth]{etavsA2.png}
%      \caption{$\delta \eta$ as a function of the Anderson-Darling coefficient $A^2$. Colors and errorbars are as in Figure \ref{fig:Msigma}.
%              }
%         \label{fig:AD}
%\end{figure}
No strong correlations can be highlighted among $A^2$ and $\delta\eta$, but dividing again the sample into two subsamples according to the median value $A^2 = 0.44$, the average  logarithmic differences in the two bins, computed by 1000 bootstrap resampling, are $\langle|\delta\eta|\rangle = 0.23\pm 0.08$ and $0.13\pm 0.07$ for $A^2 > 0.44$ and $ A^2 < 0.44 $, respectively. This small difference in mean $|\delta \eta|$ shows a weak preference of lower biases for clusters with a more regular LOS velocity dispersion. 

The \textsc{MG-MAMPOSSt} method may be sensitive to the presence of relevant substructures in clusters, which are connected to recent
merging activity and thus indicate a possible disturbed dynamical state (e.g., \citealt{Wen_2013,Parekh2015,Kimmig_2023}). Thus, we search for possible correlations between the number of substructures in the cluster and the observed logarithmic difference $\delta\eta$. We estimate the fraction of galaxies $f_\text{sub}$ in substructures over the total number of galaxies within $R=r_{200}$ by means of the DS+ method\footnote{The code is freely available at \href{https://github.com/josegit88/MilaDS}{https://github.com/josegit88/MilaDS}.} of \cite{Benavides23}. This DS+ method extends the original algorithm of \cite{Dressler88} for the identification of sub-groups in clusters by looking for galaxy clumps whose mean velocities and dispersion deviate from the global cluster values. DS+ employs a combination of projected position and line-of-sight velocities to identify substructures and estimate the probability of each galaxy
of belonging to each of those groups. 
\begin{figure}

\includegraphics[width=0.8\columnwidth]{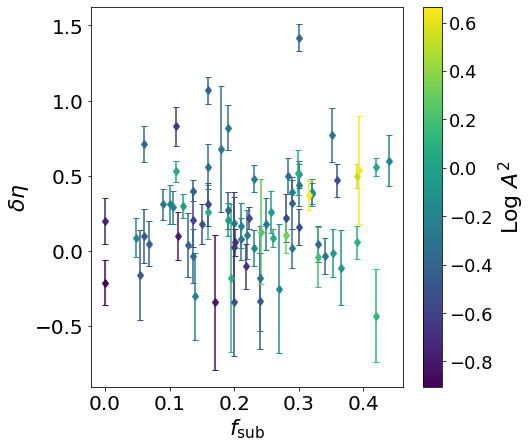}
      \caption{$\delta \eta = \text{Log}\, M_{500}^{M}- \text{Log}\, M_{500}^{SZ}$ as a function of the the fraction of galaxies in substructures $f_\text{sub}$, computed using the DS+ code of \cite{Benavides23}. Colors and errorbars are as in Figure \ref{fig:z}.
              }
         \label{fig:fg}
\end{figure}

\begin{figure}

\includegraphics[width=0.75\columnwidth]{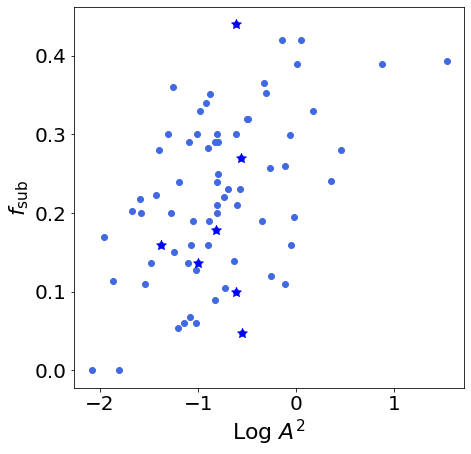}
      \caption{Fraction of galaxies in substructures $f_\text{sub}$ as a function of the logarithm of the Anderson-Darling coefficient $A^2$. The darker stars indicate the clusters for which $\text{Log}\,(M_{200}^M/M_{\sigma,200c}) > 0.3$}
         \label{fig:Avsfg}
\end{figure}

For each cluster, we run 1000
Monte Carlo simulations and we set a  p-value of $0.01$ as threshold for the identification of a substructure (see also \citealt{Barrena24}). We work in the so called non-overlapping mode, i.e., each galaxy is attributed to at most one group. We then define $f_\text{sub}= N_\text{sub}/N_{200}$, where $N_\text{sub}$ is the total number of galaxies in all the substructures found by DS+, and $N_{200}$ is the number of members in the projected cylinder $R<r_{200}$. In Figure \ref{fig:fg}, the distribution of $\delta\eta$ as a function of $f_\text{sub}$ is shown, color-coded according to the values of $A^2$. Again, no relevant trends can be drawn from the sample, with a more scattered distribution across the range of $f_\text{sub}$. 
It is worth pointing out that a $\delta \eta$ close to zero does not indicate that the cluster mass estimations are unbiased, but it may suggest that both SZ and kinematic masses are biased towards the same direction. 
Figure \ref{fig:Avsfg} further shows the relation between $f_\text{sub}$ and $\text{Log}\,A^2$. It is evident that, despite the correlation not being strong (Pearson coefficient $r = 0.43$), a trend of having large $A^2$ for large values of $f_\text{sub}$ is present.

A more accurate analysis would involve masses estimated from weak lensing (WL), as well as the combination of different morphological parameters (e.g., those derived from X-ray analyses, see \citealt{Campitiello22}) to have a clearer indication of the equilibrium state of the clusters. This will be the subject of a follow-up work.  

\section{Orbits of galaxies as a function of mass and redshift}
\label{sec:anis}
The orbital anisotropy of member galaxies in clusters is one of the key quantities in kinematic analyses. It has been shown that the nature of anisotropy can be linked to the dynamical state of the cluster and its formation history (e.g., \citealt{Mamon19}); in general, studies of observed clusters and cluster-size halos in cosmological simulations have identified an overall trend of isotropic orbits in the cluster core, and an increasing radial anisotropy towards the outskirts. Moreover, galaxies of different morphological types in clusters exhibit different orbital behaviors. In particular, early-type galaxies which are expected to have fallen into the cluster at early times may prefer more isotropic orbits, while spiral galaxies are thought to follow fairly radial paths (e.g., \citealt{Mun14,Mamon19,Biv24}).
The reconstruction of anisotropy profiles in clusters is generally a complicated issue; this is because $\beta(r)$ is not a direct observable and it is degenerate with the total mass (\citealt{Binney1982}), constituting a possible source of systematic error in the reconstruction of dynamical mass profiles. To address this, both parametric approaches (e.g., \citealt{mamon01,Read21}) and non-parametric methods (see e.g., \citealt{Mamon19} and references therein) have been developed. 
The \textsc{MG-MAMPOSSt} technique (as does the original \textsc{MAMPOSSt}) relies on a joint parametric modeling of both mass and anisotropy profiles, which has been shown to work robustly and adequately well in a wide range of cases (\citealt{mamon01,Pizzuti19b,Biviano:2023oyf}). However, the uncertainties in the reconstructed  radial profile $\beta(r)$ are quite large for a single cluster, even for a considerably large number of members in the \textsc{MG-MAMPOSSt} fit (see the example of PSZ2~G067.17+67.46 in the bottom plot of Figure \ref{fig:exampleAnis}, with $N_\text{gal} =  299$ galaxies used in the analysis).\\
This issue can be solved in two ways. One possibility is to assume prior constraints on the mass profile, as provided, for example, by hydrostatic equilibrium or WL analyses (e.g., \citealt{Annunziatella+16,AADDV17}). On the other hand, one can infer information on the anisotropy profile as a function of cluster mass and redshift by populating the PPS with a stacking procedure.
Here, we adopt the latter approach and perform two different stacking procedures. First, we identify clusters with similar velocity dispersions in two redshift bins, namely $z\le 0.2$ and $0.2 < z < 0.6$ (which mostly splits clusters into the T1 or the T2 subsample). In the first bin, we further divide clusters in three sub-bins of velocity dispersion, centered on $\sigma_\text{ap,200}/[\text{km s$^{-1}$}] = 750,\, 980,\, 1360$, respectively. We then generate the three stacked clusters (one for each bin of $\sigma_\text{ap,200}$) by considering all the member galaxies of CHEX-MATE clusters belonging to the same bin. The choices of the bins and the construction ensure that each stack contains $N\sim 3000$ galaxies. For the high-redshift bin, we end up with only two stacked clusters, with $\sigma_\text{ap,200}/[\text{km s$^{-1}$}] = 1040, 1600$. \\
The second stacking has been performed considering five bins for clusters with similar SZ-X-ray calibrated masses - three for $z < 0.2$, $M_{500}^{SZ}/10^{14}$ M$_\odot = 3.23,\,5.48,\,7.74$, and two for $z > 0.2$, $M_{500}^{SZ}/10^{14}$ M$_\odot = 7.91,\, 12.10$. This second set is to consider a stacking criterion ($M_{500}^{SZ}$) which is independent from kinematic variables. As an example, the stacked PPS for this case are shown in Appendix \ref{app:stacking}.

\begin{figure*}
   \centering
   \includegraphics[width=0.85\textwidth]{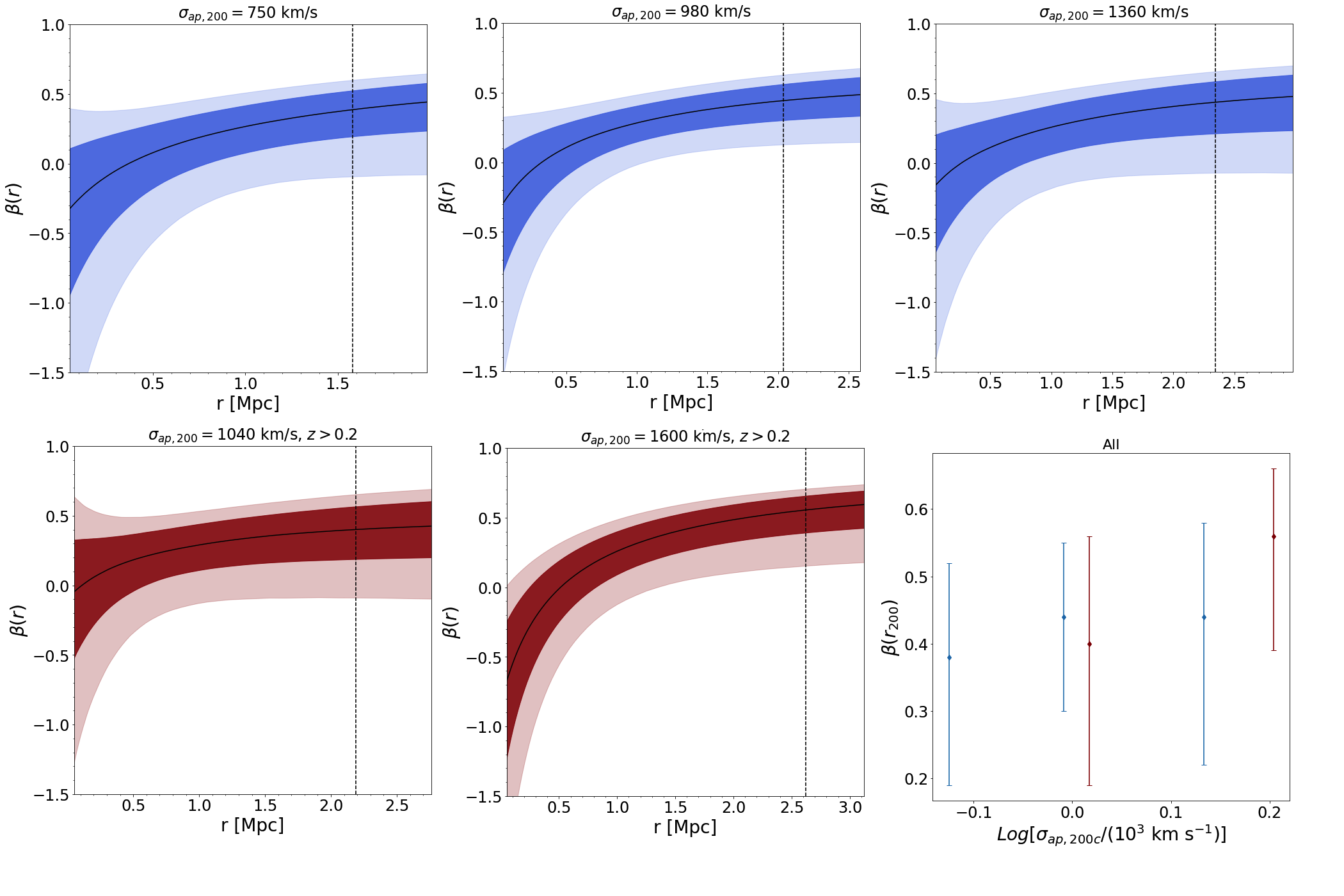}
      \caption{Radial velocity anisotropy profiles for  stacked clusters, where the stacking is in velocity dispersion. The bands show the 68\% and 95\% regions, while the black curve is the median profile. Top: $z<0.2$. Bottom: $z>0.2$. The bottom right plot shows the value of $\beta(r=r_{200})$ with 68\% errorbars (blue and red for $z<0.2$ and $z>0.2$, respectively). 
              }
         \label{fig:anis}
   \end{figure*}

\begin{figure*}
   \centering
   \includegraphics[width=0.95\textwidth]{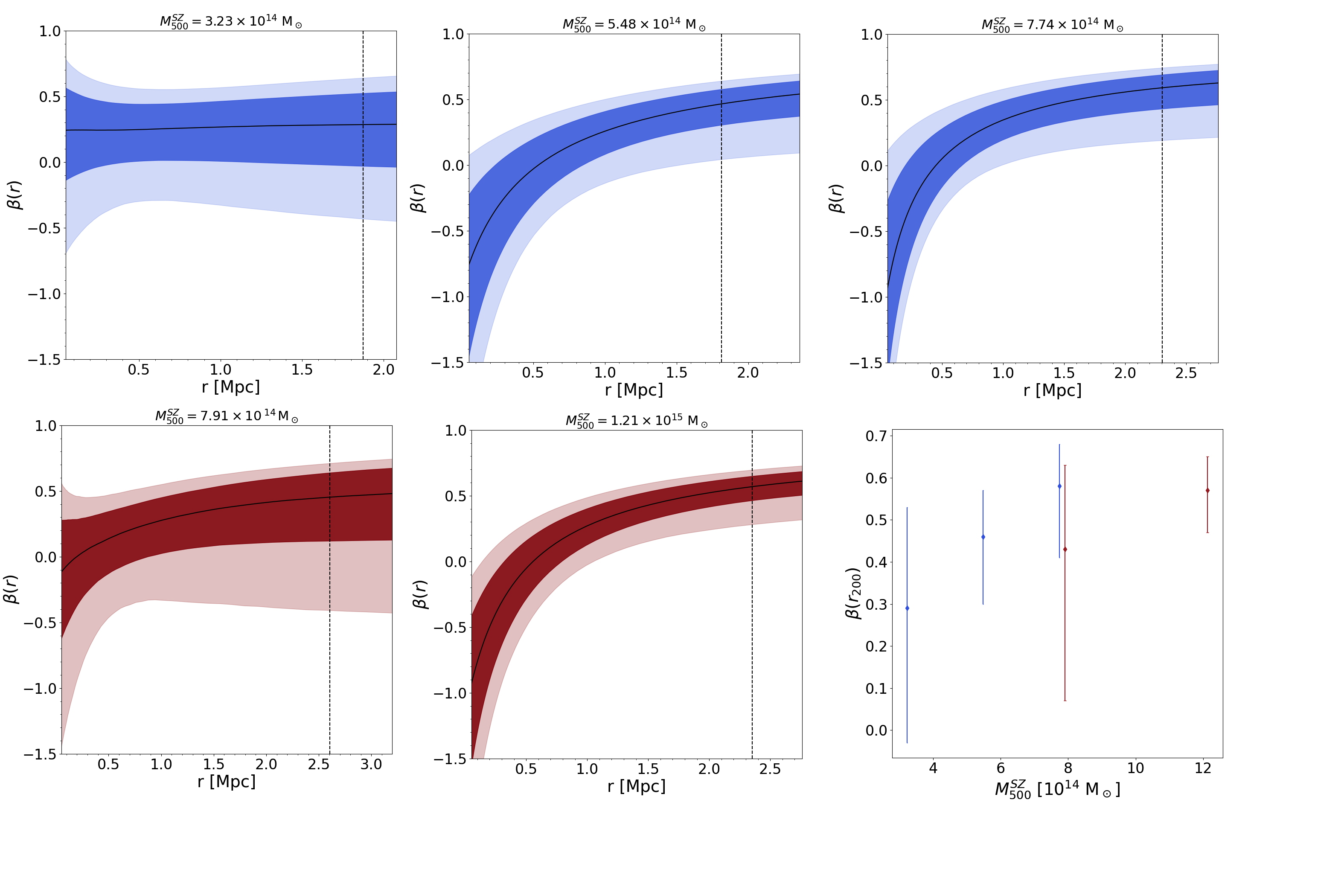}
      \caption{Radial velocity anisotropy profiles for the stacked clusters, where the stacking is in bins of $M_{500}^{SZ}$. The structure is exactly as in Fig.~\ref{fig:anis}. 
              }
         \label{fig:anisSZ}
   \end{figure*}

For each of the 10 stacked clusters, we consider galaxies in the projected range $[0.05\,\text{Mpc}, 1.1\,\langle r_\text{200,ap}\rangle]$, where $\langle r_\text{200,ap}\rangle$ is the average of $r_\text{200,ap}$ over all the clusters in the bin. We apply the \textsc{MG-MAMPOSSt} method assuming a NFW model for the mass distribution and a pNFW for the projected number density distribution; as before, we perform a MCMC sampling of the parameter space $\{r_{200},\,r_\text{s},r_\nu,\,\mathcal{A}_0,\,\mathcal{A}_\infty\}$, adopting the same flat prior listed in Sect.~\ref{sec:setup}.  \\
Fig.~\ref{fig:anis}  and Fig.~\ref{fig:anisSZ} show the radial profiles of the velocity anisotropy obtained from the \textsc{MG-MAMPOSSt} analysis of the stacked clusters ($\sigma_\text{ap,200}$ and SZ, respectively) at $z<0.2$ (top panels in each plot) and $z>0.2$ (bottom panels in each plot). 
For the stacking in velocity dispersion, we see that all the profiles (except the one for the most massive cluster, $\sigma_\text{ap,200} \sim 1600$ km s$^{-1}$) are consistent with central isotropic orbits within one $\sigma$, and with an increase of the anisotropy towards the cluster outskirts, as found by other studies in the literature \citep{Mamon19}. 

In the bottom right panel, we can observe a slight, although not very significant, trend to have more radial orbits at the virial radius for clusters with a larger velocity dispersion. The most massive stacked cluster exhibits a more 'extreme' behavior, passing from tangential orbits ($\beta < 0$) at the center, to largely radial ones in the outskirts. However, this bin contains clusters with $\delta\eta > 0.8$, which, as we have discussed, are characterized by disturbed dynamical states.  

The same trend of increasing radial orbits at virial radius is found in terms of the SZ mass, as highlighted in Fig.~\ref{fig:anisSZ}. In this case, however, only the first mass sub-bins in each redshift cut ($z<0.2$ and $z>0.2$) exhibit a central anisotropy consistent with zero at the 68\% C.L., while the others are in agreement with no anisotropy at the 95\% C.L., with a slight preference for central tangential orbits which is not very statistically significant. 

%Results at r200: $\beta(r_{200},750) = %0.38^{+0.14}_{-0.19}$,  
%$\beta(r_{200},980) = 0.44^{+0.11}_{-0.14}$,  
%$\beta(r_{200},1360) = 0.44^{+0.14}_{-0.22}$, 
%$\beta(r_{200},1040,z>0.2) = 0.40^{+0.16}_{-0.21}$, 
%$\beta(r_{200},1600,z>0.2) = 0.56^{+0.10}_{-0.17}$, 
%                                     Two column Table
%______________________________________________ 


%                                     Two column figure (place early!)
%______________________________________________ 
%______________________________________________________________

%36 NFW 8 Her 26 Bur 10 Eis Dire sopra che NFW è comunque sempre fit adeguato

\section{Conclusions}
In this paper we have studied the kinematical and dynamical properties of 75 massive galaxy clusters, as part of the CHEX-MATE sample, for which high-precision spectroscopic measurements of LOS velocities and projected positions are available. By means of the \textsc{MG-MAMPOSSt} code, we have jointly reconstructed mass profiles and velocity anisotropy profiles for eight combinations of mass and number density models. We found that the NFW prescription is preferred by the kinematic analysis for about half of the sample, followed by the Burkert model. This confirms the assumptions of the halo model used in the dynamical mass determinations of \cite{Sereno24}.
Moreover, the constraints on the virial radius $r_{200}$ (and on the mass $M_{200}$) are robust against variations in the mass and number density models, as shown by the relatively small systematic uncertainties (Table \ref{tab:results}). 

We have compared the masses obtained from the SZ-X-ray calibrated signal, finding $(1-B_1) = 0.54\pm 0.11$ when clusters with evident disturbed dynamical states are removed; a negligible bias is instead obtained while fitting with a linear relation allowing for mass dependence, $(1-B) = 1.04\pm 0.09$. When also the clusters with a logarithmic mass difference $\delta \eta >0.8$ dex (which are high-velocity dispersions $\sigma_\text{ap,200} > 1.4\times10^{3}$ km/s) are considered,  an overall positive mass-dependent bias $(1-B) = 0.67$ is found. Note that SZ masses may be overestimated due to the Malmquist bias, as discussed in \cite{Sereno24}; this might affect our findings producing a larger value of $(1-B)$. A refined analysis is however beyond the scope of the current work, as it has already been performed in \cite{Sereno24}.
%For 54 clusters, the Weak Lensing (WL) mass estimates from the CoMaLit Litearture Catalog (LC2) of \cite{Sereno2015} have been further considered. Compared to the SZ case, the relation among the \textsc{MG-MAMPOSSt} and WL measurements of $M_{500}$ exhibits a  wider scatter with a flatter slope and a quite large bias $(1-B) = 2.2$. However, when forcing the slope to $1$, the same bias $(1-B) = 0.67$ as for the SZ case is obtained.
The concentration-mass relation has been further investigated and compared to theoretical expectations. By performing a Bayesian linear regression by means of the LIRA package, we found a slight evidence of a rising trend at large masses, as also predicted by the model of \cite{Diemer19}. Interestingly, the results obtained for the sample of CHEX-MATE clusters are fully consistent with the prediction of the $\Lambda$CDM scenario, given an average value of $\langle c_{200} \rangle = 3.5$ with a scatter of 1.8.
The results of the various scaling relations investigated in this work are summarized in Table \ref{tab:correlations}.
\begin{table*}
\centering
\caption{\label{tab:correlations} Summary of the results for the parameters of the linear fits of the relations discussed in this paper.}
\begin{tabular}{c|c|c}
\toprule
\hline
Relation & reference  & constraints \\
\midrule

$\text{Log}\,M_{200}^{M}-\sigma_\text{ap,200}$ & Eq. \eqref{eq:fitmsigma} & $\alpha_\sigma = 3.61 \pm 0.17,\, \beta_\sigma = 1.07 \pm 0.01, \,\gamma_\sigma =-0.83 \pm 0.70$   \\[0.2 cm]
$\text{Log}\,M_{500}^{M}-\text{Log}\,M_{500}^{SZ}$ (outliers removed) & Eq. \eqref{eq:logmfit} & $\alpha_{SZ} = 0.71 \pm 0.04,\, 1-B = 1.04 \pm 0.09$ \\
& $\alpha_{SZ}$ fixed to 1 & $(1-B_1) = 0.54 \pm 0.11$\\[0.2cm]
$\text{Log}\, c_{200}- \text{Log}\, M^{M}_{200}$ (outliers removed) & Eq. \eqref{eq_scaling_1} & $\alpha_C = 0.49 \pm 0.07,\, \beta_C = 0.20 \pm 0.31, \, \gamma_C = - 0.23 \pm 1.6$ \\   
\bottomrule
\end{tabular}
\end{table*}

The dynamical state of the clusters has been studied by comparing $\delta \eta$ with two quantities that can be related with dynamical un-relaxation and disturbed morphology, namely the Anderson-Darling coefficient $A^2$ and the fraction of galaxies in substructures $f_\text{sub}$. No evident relations have been found for $A^2$ and $f_\text{sub}$. Nevertheless, a lower averaged $\delta \eta$ is obtained for clusters with $A^2<0.44$ (being 0.44 the median value of the sample).
Note however that a more complete analysis should investigate the differences with the weak lensing mass determinations as well as account for the X-ray morphology to better identify substructures and departures from equilibrium. This will be the subject of an upcoming work.

Finally, the shape of the radial velocity anisotropy profile has been studied by dividing the sample in two bins of redshift, and first collecting clusters within three and two sub-bins of $\sigma_\text{ap,200}$, respectively. We stacked clusters in the same bin and applied \textsc{MG-MAMPOSSt} to those stacked systems to reconstruct the orbits from the center to the edges. We performed a similar stacking procedure considering bins of $M_{500}^{SZ}$, which are not directly related to the kinematical properties of the clusters.
Consistent with the literature, orbits tend to be isotropic or tangential at the center and more radial towards the edge of the clusters, where galaxies are thought to be falling in the cluster's potential for the first time \citep{Mamon19}. There is a slight trend towards increasingly radial orbits at $r_{200}$ with larger $\sigma_\text{ap,200}$ (which, in turn, is connected to larger dynamical masses); we found the same trend with increasing $M_{500}^{SZ}$.\\
More broadly, our results indicate that the most massive galaxy clusters at 
$z < 0.6$ are generally close to dynamical equilibrium. Nonetheless, we identify some cases where even the most massive systems appear to deviate from full virialisation. As previously mentioned, a comprehensive assessment of the dynamical state of the CHEX-MATE sample will be carried out in a forthcoming study, which will incorporate a more complete set of morphological and kinematic indicators, as well as a comparison with mass estimates from weak lensing and X-ray data.

\begin{acknowledgements}
This project has been partially funded by the Spanish Ministerio de Ciencia, Innovación y Universidades through the programme Generación del Concimiento 2021, with code PID2021-122665OB-I00. R. Barrena acknowledges support by the Severo Ochoa 2020 research programme of the Instituto de Astrofísica de Canarias. 
S.E., M.S., acknowledge the financial contribution from the contracts Prin-MUR 2022 supported by Next Generation EU (M4.C2.1.1, n.20227RNLY3 {\it The concordance cosmological model: stress-tests with galaxy clusters}).
M.S. acknowledges financial contributions from INAF Theory Grant 2023: Gravitational lensing detection of matter distribution at galaxy cluster boundaries and beyond (1.05.23.06.17).
S.E. acknowledges the financial contribution from the European Union’s Horizon 2020 Programme under the AHEAD2020 project (grant agreement n. 871158).
DE acknowledges support from the Swiss National Science Foundation (SNSF) through grant agreement 200021\_212576. C.P.H. acknowledges support from ANID through Fondecyt Regular project number 1252233. JS was supported by NASA Astrophysics Data Analysis Program (ADAP) Grant 80NSSC21K1571. BJM acknowledges support from Science and Technology Facilities Council grants ST/V000454/1 and ST/Y002008/1. GWP acknowledges long-term support from CNES, the French space agency. GC  acknowledges the support from the Next Generation EU funds within the National Recovery and Resilience Plan (PNRR), Mission 4 - Education and Research, Component 2 - From Research to Business (M4C2), Investment Line 3.1 - Strengthening and creation of Research Infrastructures, Project IR0000012 – “CTA+ - Cherenkov Telescope Array Plus”. MD acknowledges the support of two NASA programs: NASA award 80NSSC19K0116/  SAO subaward SV9-89010 and NASA award 80NSSC 22K0476.

\end{acknowledgements}

%-------------------------------------------------------------------

\bibliographystyle{aa}
\bibliography{sample631}

\begin{appendix}  % Start the appendix section

\section{Results from \textsc{MG-MAMPOSSt}}  % First appendix section
\label{app:results}

\setlength{\tabcolsep}{2.5pt}
The constraints on the mass profile, number density and velocity anisotropy parameters are summarized in Table \ref{tab:results}, along with cluster redshift, Anderson-Darling coefficient $A^2$, fraction of galaxies in substructures $f_\text{sub}$, number of galaxies $N_{200}$ within a projected cylinder $R < r_\text{ap,200}$ aperture velocity dispersion. Here we included also the \textsc{MG-MAMPOSSt} analysis of clusters PSZ2G113.91-37.01 and PSZ2G083.29-31.03, which have less than $N=50$ members within $r_\text{200,ap}$.
\begin{sidewaystable*}
%\centering
\caption{\label{tab:results} Constraints on the free parameters from the kinematic analysis with \textsc{MG-MAMPOSSt}, shown for the combination of mass and number density model which provides the best BIC. }

\begin{tabular}{c|c|c|c|c|c|c|c|c|c|c|c|c|c}

\toprule
\hline
PSZ2 name & Mass  & $N(R)$ & $r_{200}$ & $r_{\nu}$ & $r_\text{s}$ & $\mathcal{A}_\infty$ & $\mathcal{A}_0$ & $-\ln P$ & $z$ & $A^2$ & $f_\text{sub}$ & $N_{200}$ & $\sigma_\text{ap,200}$  \\
   &       &        &  [Mpc]   &  [Mpc]   & [Mpc]  & & & & & &           &   & $[\text{km/s}] $ \\ 
\midrule
  G049.22+30.87 & NFW & pNFW & $2.13^{+0.25 }_{-0.27} \pm 0.06$ & $0.59^{+0.10 }_{-0.13} \pm 0.28$ & $0.81^{+1.22 }_{-0.72} \pm 0.73$ & $1.42^{+1.76 }_{-0.93} \pm 0.09$ & $1.78^{+2.36 }_{-1.33} \pm 0.10$ &  3403.919 & 0.1606 & 0.27 & $ 0.29 $ & 368 & 1034.0  \\ [0.2cm]
  G226.18+76.79 & Bur & pHer & $2.29^{+0.53 }_{-0.44} \pm 0.04$ & $0.67^{+0.17 }_{-0.13} \pm 0.24$ & $0.30^{+0.41 }_{-0.28} \pm 0.38$ & $1.64^{+1.94 }_{-1.18} \pm 0.04$ & $1.46^{+2.71 }_{-1.00} \pm 0.14$ & 1171.266 & 0.1415 & 0.44 & $0.09$ & 131 & 1069.0  \\ [0.2cm]
  G053.53+59.52 & NFW & pHer & $2.40^{+0.36 }_{-0.32} \pm 0.08$ & $0.87^{+0.14 }_{-0.15} \pm 0.30$ & $0.85^{+1.21 }_{-0.73} \pm 0.66$ & $1.58^{+1.90 }_{-1.10} \pm 0.12$ & $1.33^{+2.16 }_{-0.89} \pm 0.16$ & 2100.914 & 0.1136 & 0.78 & 0.12 & 222& 1056.0  \\ [0.2cm]
    G056.77+36.32  & NFW & pNFW & $3.41^{+0.81 }_{-0.76} \pm 0.15$ & $0.85^{+0.29 }_{-0.16} \pm 0.74$ & $2.09^{+1.64 }_{-1.54} \pm 0.46$ & $1.38^{+2.14 }_{-0.91} \pm 0.14$ & $1.17^{+2.42 }_{-0.70} \pm 0.39$ &  981.519 & 0.09696 & 0.41 & 0.19 & 96 &	1364.0  \\ [0.2cm]
     G042.81+56.61  & NFW & pNFW & $2.48^{+0.31 }_{-0.29} \pm 0.08$ & $0.53^{+0.17 }_{-0.15} \pm 0.49$ & $0.63^{+0.76 }_{-0.52} \pm 0.77$ & $1.63^{+1.78 }_{-1.06} \pm 0.08$ & $1.44^{+2.38 }_{-1.00} \pm 0.10$ &  2014.210 & 0.07225 & 0.54 & 0.30 & 210 & 1168.0  \\ [0.2cm]
    G033.81+77.18  & Bur & pNFW & $1.79^{+0.31 }_{-0.25} \pm 0.05$ & $0.34^{+0.12 }_{-0.12} \pm 0.34$ & $0.40^{+0.56 }_{-0.37} \pm 0.61$ & $1.48^{+1.82 }_{-1.01} \pm 0.08$ & $2.09^{+2.38 }_{-1.60} \pm 0.17$ &  1293.172 & 0.06272 & 0.56 & 0.23 & 154 & 818.0  \\ [0.2cm]
      G041.45+29.10  & Bur & pHer & $2.40^{+0.52 }_{-0.44} \pm 0.15$ & $0.38^{+0.10 }_{-0.08} \pm 0.14$ & $0.12^{+0.15 }_{-0.08} \pm 0.15$ & $1.74^{+1.78 }_{-1.22} \pm 0.09$ & $1.77^{+2.46 }_{-1.32} \pm 0.10$ & 961.440 & 0.1771 & 0.28 & 0.37 &	110 & 1332.0  \\ [0.2cm]
    G238.69+63.26  & Her & pHer & $1.83^{+0.28 }_{-0.26} \pm 0.07$ & $1.95^{+0.20 }_{-0.29} \pm 0.52$ & $1.56^{+1.74 }_{-1.30} \pm 0.46$ & $1.51^{+2.00 }_{-1.05} \pm 0.08$ & $1.22^{+2.32 }_{-0.76} \pm 0.16$ & 1770.555 & 0.1672 & 0.45 & 0.21 & 229	& 897.1  \\ [0.2cm]
   G080.41-33.24  & NFW & pHer & $2.13^{+0.62 }_{-0.55} \pm 0.18$ & $0.36^{+0.08 }_{-0.08} \pm 0.14$ & $0.99^{+1.81 }_{-0.97} \pm 0.69$ & $1.34^{+1.74 }_{-0.89} \pm 0.11$ & $2.13^{+2.34 }_{-1.62} \pm 0.09$ & 1124.055 & 0.1101 & 0.54 & 0.10 & 137 &	 801.8  \\ [0.2cm]
  G048.10+57.16  & Bur & pHer & $1.87^{+0.36 }_{-0.33} \pm 0.06$ & $1.11^{+0.24 }_{-0.15} \pm 0.50$ & $0.64^{+1.10 }_{-0.63} \pm 0.73$ & $1.49^{+1.84 }_{-1.03} \pm 0.09$ & $2.00^{+2.38 }_{-1.54} \pm 0.10$ & 1546.255 & 0.07782 & 0.48 & 0.22 & 165 & 823.7  \\ [0.2cm]
  G287.46+81.12  & Eis & pHer & $1.69^{+0.31 }_{-0.29} \pm 0.06$ & $1.72^{+0.36 }_{-0.46} \pm 0.50$ & $0.64^{+1.47 }_{-0.64} \pm 0.73$ & $1.85^{+1.83 }_{-1.32} \pm 0.09$ & $1.96^{+2.42 }_{-1.50} \pm 0.10$ & 964.668 & 0.07194 & 0.25 & 0.28 & 109 & 832.7  \\ [0.2cm]
  G031.93+78.71  & Bur & pNFW & $1.28^{+0.56 }_{-0.73} \pm 0.06$ & $0.35^{+0.31 }_{-0.18} \pm 0.50$ & $0.82^{+1.33 }_{-0.82} \pm 0.73$ & $1.76^{+1.85 }_{-1.27} \pm 0.09$ & $2.14^{+2.35 }_{-1.63} \pm 0.10$ & 400.254 & 0.07533 & 0.28 & 0.20 & 57	& 589.5  \\ [0.2cm]
  G067.17+67.46  & NFW & pNFW & $2.19^{+0.28 }_{-0.29} \pm 0.06$ & $0.74^{+0.20 }_{-0.19} \pm 0.50$ & $0.85^{+1.34 }_{-0.75} \pm 0.73$ & $1.55^{+1.72 }_{-1.01} \pm 0.09$ & $1.65^{+2.56 }_{-1.20} \pm 0.10$ & 2880.857 & 0.1667 & 1.57 & 0.28 & 291 & 1057.0  \\ [0.2cm]
  G006.49+50.56  & NFW & pNFW & $2.13^{+0.16 }_{-0.15} \pm 0.06$ & $1.08^{+0.19 }_{-0.14} \pm 0.50$ & $0.40^{+0.35 }_{-0.27} \pm 0.73$ & $2.02^{+1.58 }_{-1.08} \pm 0.09$ & $1.04^{+1.19 }_{-0.58} \pm 0.10$ & 6711.942 & 0.07854 & 0.89 & 0.26 & 750 & 1066.0  \\ [0.2cm]
  %SONO QUIIIIII =======================================================
  G124.20-36.48  & Her & pHer & $2.61^{+0.39 }_{-0.37} \pm 0.06$ & $1.22^{+0.11 }_{-0.15} \pm 0.50$ & $1.23^{+1.65 }_{-1.10} \pm 0.73$ & $1.55^{+1.86 }_{-1.08} \pm 0.09$ & $1.71^{+2.45 }_{-1.26} \pm 0.10$ & 2031.290 & 0.1921 & 0.45 & 0.35	& 232 &  1282.0  \\ [0.2cm]
  G021.10+33.24  & Eis & pNFW & $2.34^{+0.52 }_{-0.75} \pm 0.06$ & $1.79^{+1.38 }_{-0.98} \pm 0.50$ & $0.70^{+2.01 }_{-0.68} \pm 0.73$ & $1.59^{+2.05 }_{-1.12} \pm 0.09$ & $2.08^{+2.42 }_{-1.63} \pm 0.10$ & 643.999 & 0.1511 & 0.95 & 0.08 & 74	& 1296.0  \\ [0.2cm]
  G044.20+48.66  & NFW & pHer & $2.38^{+0.26 }_{-0.28} \pm 0.06$ & $2.33^{+0.14 }_{-0.21} \pm 0.50$ & $1.60^{+1.59 }_{-1.20} \pm 0.73$ & $1.42^{+1.74 }_{-0.88} \pm 0.09$ & $1.53^{+2.16 }_{-0.99} \pm 0.10$ & 7619.565 & 0.08987 & 0.44 & 0.33 & 778 & 1090.0  \\ [0.2cm]
  G075.71+13.51  & Her & pHer & $3.97^{+0.88 }_{-0.85} \pm 0.06$ & $0.50^{+0.13 }_{-0.11} \pm 0.50$ & $1.64^{+1.47 }_{-1.18} \pm 0.73$ & $1.26^{+1.95 }_{-0.79} \pm 0.09$ & $1.63^{+2.47 }_{-1.19} \pm 0.10$ & 1249.919 & 0.05455 & 0.21 & 0.12  & 134 & 1723.0  \\ [0.2cm]
  G324.04+48.79  & Eis & pHer & $1.79^{+0.48 }_{-0.42} \pm 0.06$ & $0.92^{+0.42 }_{-0.29} \pm 0.50$ & $0.97^{+2.04 }_{-0.98} \pm 0.73$ & $1.61^{+1.95 }_{-1.14} \pm 0.09$ & $1.96^{+2.33 }_{-1.51} \pm 0.10$ & 634.364 & 0.4505 & 0.12 & 0.00 & 78	& 939.7  \\ [0.2cm]

   G284.41+52.45   & NFW & pHer & $2.07^{+0.23 }_{-0.26} \pm 0.04$ & $2.19^{+0.40 }_{-0.45} \pm 0.24$ & $0.90^{+1.56 }_{-0.89} \pm 0.38$ & $1.49^{+1.52 }_{-0.85} \pm 0.04$ & $2.08^{+2.27 }_{-1.52} \pm 0.14$ & 2983.187 & 0.4399 & 0.40 & 0.33 & 333 & 1247.0  \\ [0.2cm]
  G349.46-59.95   & NFW & pHer & $3.46^{+1.16 }_{-1.09} \pm 0.04$ & $0.62^{+0.27 }_{-0.22} \pm 0.24$ & $1.35^{+1.92 }_{-1.34} \pm 0.38$ & $1.31^{+1.87 }_{-0.85} \pm 0.04$ & $2.35^{+2.27 }_{-1.71} \pm 0.14$ & 719.670 & 0.3465 & 0.54	& 0.19 & 81 & 1591.0  \\ [0.2cm]
  G008.94-81.22  & Bur & pHer & $3.05^{+0.32 }_{-0.30} \pm 0.04$ & $0.90^{+0.09 }_{-0.06} \pm 0.24$ & $0.26^{+0.18 }_{-0.17} \pm 0.38$ & $2.24^{+1.62 }_{-1.41} \pm 0.04$ & $1.17^{+1.96 }_{-0.73} \pm 0.14$ &  4631.269 & 0.3064 & 0.87 & 0.39 & 491 & 1763.0  \\ [0.2cm]
  G266.04-21.25   & NFW & pHer & $2.98^{+0.92 }_{-0.83} \pm 0.04$ & $0.46^{+0.10 }_{-0.12} \pm 0.24$ & $1.25^{+1.90 }_{-1.21} \pm 0.38$ & $1.43^{+1.85 }_{-0.98} \pm 0.04$ & $1.95^{+2.35 }_{-1.50} \pm 0.14$ & 838.472 & 0.2965 & 0.25 & 0.15 & 93	& 1275.0  \\ [0.2cm]
%   G262.27-35.38  & Bur & pHer & $1.54^{+0.73 }_{-0.63} \pm 0.04$ & $0.78^{+0.42 }_{-0.32} \pm 0.24$ & $0.67^{+1.80 }_{-0.65} \pm 0.38$ & $1.62^{+1.88 }_{-1.16} \pm 0.04$ & $1.85^{+2.40 }_{-1.40} \pm 0.14$ & 405.296 & 0.2937 & 0.2564 & 727.5  \\ [0.2cm]
   \bottomrule
\end{tabular}

\end{sidewaystable*}

\begin{sidewaystable*}
\begin{tabular}{c|c|c|c|c|c|c|c|c|c|c|c|c|c}
\toprule
\hline
PSZ2 name & Mass  & $N(R)$ & $r_{200}$ & $r_{\nu}$ & $r_\text{s}$ & $\mathcal{A}_\infty$ & $\mathcal{A}_0$ & $-\ln P$ & $z$ & $A^2$ & $f_\text{sub}$ & $N_{200}$ & $\sigma_\text{ap,200}$  \\
   &       &        &  [Mpc]   &  [Mpc]   & [Mpc]  & & & & & &           &   & $[\text{km/s}] $ \\ 
\midrule
 G107.10+65.32  & Bur & pHer & $3.00^{+0.34 }_{-0.32} \pm 0.04$ & $1.74^{+0.10 }_{-0.15} \pm 0.24$ & $0.69^{+0.67 }_{-0.51} \pm 0.38$ & $1.62^{+1.92 }_{-1.14} \pm 0.04$ & $1.38^{+2.33 }_{-0.93} \pm 0.14$ & 3463.596 & 0.2772 & 1.05 & 0.41 & 348 & 1491.0  \\ [0.2cm]
 G340.36+60.58  & Eis & pHer & $2.84^{+0.42 }_{-0.49} \pm 0.04$ & $3.56^{+0.62 }_{-0.73} \pm 0.24$ & $1.53^{+1.77 }_{-1.17} \pm 0.38$ & $1.16^{+1.73 }_{-0.69} \pm 0.04$ & $1.72^{+2.47 }_{-1.26} \pm 0.14$ & 3070.106 & 0.251 & 2.38 & 0.38 & 368 & 1426.0  \\ [0.2cm]
 G072.62+41.46  & NFW & pHer & $2.52^{+0.33 }_{-0.33} \pm 0.04$ & $2.13^{+0.42 }_{-0.34} \pm 0.24$ & $1.54^{+1.81 }_{-1.39} \pm 0.38$ & $1.23^{+1.55 }_{-0.72} \pm 0.04$ & $2.11^{+2.30 }_{-1.44} \pm 0.14$ & 3316.267 & 0.2255 & 0.34 & 0.19 & 368	& 1260.0  \\ [0.2cm]
  G055.59+31.85  & NFW & pHer & $2.03^{+0.51 }_{-0.50} \pm 0.04$ & $3.74^{+1.12 }_{-1.28} \pm 0.24$ & $1.44^{+1.90 }_{-1.36} \pm 0.38$ & $1.55^{+1.94 }_{-1.09} \pm 0.04$ & $1.38^{+2.68 }_{-0.93} \pm 0.14$ & 1095.847 & 0.2255 & 0.37 & 0.21 & 126	& 1056.0  \\ [0.2cm]
 G159.91-73.50  & Eis & pHer & $2.70^{+0.31 }_{-0.29} \pm 0.08$ & $1.59^{+0.32 }_{-0.27} \pm 0.30$ & $0.74^{+0.78 }_{-0.57} \pm 0.66$ & $1.74^{+1.91 }_{-1.28} \pm 0.12$ & $1.04^{+2.12 }_{-0.56} \pm 0.16$ & 2603.277 & 0.2095 & 1.19 &	0.36 & 272 & 1370.0  \\ [0.2cm]
 G195.75-24.32  & NFW & pHer & $2.15^{+0.33 }_{-0.30} \pm 0.08$ & $1.32^{+0.24 }_{-0.29} \pm 0.30$ & $0.98^{+1.74 }_{-0.99} \pm 0.66$ & $1.49^{+1.84 }_{-1.02} \pm 0.12$ & $1.79^{+2.37 }_{-1.34} \pm 0.16$ & 1430.192 & 0.2007 & 0.60 & 0.32 & 153 & 1043.0  \\ [0.2cm]
  G313.33+61.13  & Eis & pHer & $2.87^{+0.36 }_{-0.32} \pm 0.08$ & $0.93^{+0.12 }_{-0.09} \pm 0.30$ & $0.28^{+0.20 }_{-0.18} \pm 0.66$ & $2.24^{+1.62 }_{-1.40} \pm 0.12$ & $1.16^{+1.74 }_{-0.71} \pm 0.16$ & 4277.167 & 0.1848 & 1.00 & 0.40 & 492	& 1669.0  \\ [0.2cm]
 G208.80-30.67  & Bur & pHer & $2.78^{+1.09 }_{-0.76} \pm 0.08$ & $0.83^{+0.31 }_{-0.26} \pm 0.30$ & $0.68^{+1.46 }_{-0.65} \pm 0.66$ & $1.61^{+1.93 }_{-1.15} \pm 0.12$ & $1.44^{+2.52 }_{-0.98} \pm 0.16$ & 1141.373 & 0.2468 & 0.89 & 0.27 & 130	& 1242.0  \\ [0.2cm]
 G172.98-53.55  & NFW & pHer & $3.08^{+0.77 }_{-0.67} \pm 0.08$ & $0.24^{+0.03 }_{-0.03} \pm 0.30$ & $0.39^{+0.45 }_{-0.32} \pm 0.66$ & $1.41^{+1.96 }_{-0.95} \pm 0.12$ & $1.48^{+2.47 }_{-1.03} \pm 0.16$ & 2412.888 & 0.374 & 0.44 & 0.36 & 272 & 1682.0  \\ [0.2cm]
 G205.93-39.46  & NFW & pNFW & $2.57^{+0.69 }_{-0.62} \pm 0.08$ & $0.57^{+0.46 }_{-0.28} \pm 0.30$ & $1.48^{+1.91 }_{-1.43} \pm 0.66$ & $1.33^{+1.85 }_{-0.87} \pm 0.12$ & $1.99^{+2.42 }_{-1.52} \pm 0.16$ & 544.822 & 0.4409 & 0.36	& 0.07 & 65 & 1404.0  \\ [0.2cm]
  G273.59+63.27 & Bur & pHer & $2.63^{+0.35 }_{-0.33} \pm 0.15$ & $1.74^{+0.48 }_{-0.34} \pm 0.74$ & $0.49^{+0.43 }_{-0.33} \pm 0.46$ & $2.08^{+1.71 }_{-1.42} \pm 0.14$ & $0.96^{+1.21 }_{-0.49} \pm 0.39$ & 1685.399 & 0.1335 & 0.44 & 0.19 & 180	& 1288.0  \\ [0.2cm]
  G057.61+34.93 & NFW & pHer & $2.19^{+0.35 }_{-0.32} \pm 0.15$ & $1.13^{+0.28 }_{-0.28} \pm 0.74$ & $0.62^{+1.30 }_{-0.62} \pm 0.46$ & $1.52^{+1.73 }_{-0.98} \pm 0.14$ & $1.93^{+2.41 }_{-1.47} \pm 0.39$ & 1232.084 & 0.08492 & 0.50 & 0.22 & 141 & 1021.0  \\ [0.2cm]
  G046.88+56.48 & NFW & pHer & $2.34^{+0.47 }_{-0.43} \pm 0.15$ & $2.69^{+0.80 }_{-0.54} \pm 0.74$ & $2.29^{+1.46 }_{-1.36} \pm 0.46$ & $1.60^{+1.96 }_{-1.13} \pm 0.14$ & $0.84^{+0.54 }_{-0.36} \pm 0.39$ & 1834.999 & 0.1137 & 0.61 & 0.27 & 204 & 1087.0  \\ [0.2cm]
  G243.64+67.74 & Bur & pNFW & $1.41^{+0.56 }_{-0.86} \pm 0.15$ & $0.61^{+0.43 }_{-0.25} \pm 0.74$ & $0.84^{+1.52 }_{-0.84} \pm 0.46$ & $1.61^{+1.96 }_{-1.15} \pm 0.14$ & $2.30^{+2.33 }_{-1.81} \pm 0.39$ & 488.051 & 0.08108 & 0.55 & 0.30 & 54 & 696.2  \\ [0.2cm]
  G008.31-64.74 & NFW & pHer & $4.43^{+0.57 }_{-0.61} \pm 0.15$ & $0.68^{+0.11 }_{-0.12} \pm 0.74$ & $1.39^{+1.14 }_{-0.98} \pm 0.46$ & $1.04^{+1.80 }_{-0.57} \pm 0.14$ & $1.85^{+2.43 }_{-1.33} \pm 0.39$ & 2369.679 & 0.3153 & 0.57 & 0.26 & 249 & 1978.0  \\ [0.2cm]
  G285.63+72.75 & Eis & pHer & $2.12^{+0.36 }_{-0.33} \pm 0.15$ & $1.36^{+0.33 }_{-0.36} \pm 0.74$ & $0.95^{+1.78 }_{-0.89} \pm 0.46$ & $1.58^{+1.95 }_{-1.12} \pm 0.14$ & $1.74^{+2.48 }_{-1.29} \pm 0.39$ & 1243.294 & 0.1667 & 0.34 & 0.14 & 140	& 995.7  \\ [0.2cm]
 G077.90-26.63 & Bur & pNFW & $1.86^{+0.61 }_{-0.88} \pm 0.08$ & $0.45^{+0.24 }_{-0.15} \pm 0.49$ & $0.73^{+1.60 }_{-0.73} \pm 0.77$ & $1.71^{+1.83 }_{-1.24} \pm 0.08$ & $2.29^{+2.28 }_{-1.78} \pm 0.10$ & 725.684 & 0.1449 & 0.29 & 0.14 & 81	& 915.0  \\ [0.2cm]
 G114.79-33.71 & NFW & pNFW & $2.06^{+0.37 }_{-0.34} \pm 0.08$ & $0.44^{+0.19 }_{-0.14} \pm 0.49$ & $0.87^{+1.51 }_{-0.80} \pm 0.77$ & $1.49^{+1.87 }_{-1.02} \pm 0.08$ & $1.64^{+2.49 }_{-1.19} \pm 0.10$ & 1316.109 & 0.09447 & 0.20 &	0.21 & 155 & 953.6  \\ [0.2cm]
 G224.00+69.33 & NFW & pHer & $2.28^{+0.50 }_{-0.44} \pm 0.05$ & $0.96^{+0.31 }_{-0.24} \pm 0.34$ & $0.88^{+1.96 }_{-0.88} \pm 0.61$ & $1.45^{+1.88 }_{-0.99} \pm 0.08$ & $2.06^{+2.36 }_{-1.58} \pm 0.17$ & 983.873 & 0.1917 & 0.35 & 0.20 & 114	& 1078.0  \\ [0.2cm]
 G113.91-37.01 & Eis & pHer & $2.87^{+1.07 }_{-0.96} \pm 0.05$ & $0.59^{+0.13 }_{-0.17} \pm 0.34$ & $1.15^{+1.97 }_{-1.15} \pm 0.61$ & $1.60^{+1.91 }_{-1.14} \pm 0.08$ & $2.12^{+2.38 }_{-1.65} \pm 0.17$ & 553.347 & 0.3677 & 0.57 & 0.00 & 19	& 1333.0  \\ [0.2cm]
 G067.52+34.75 & NFW & pHer & $2.03^{+0.44 }_{-0.39} \pm 0.05$ & $1.96^{+0.68 }_{-0.62} \pm 0.34$ & $1.27^{+1.87 }_{-1.24} \pm 0.61$ & $1.54^{+1.90 }_{-1.06} \pm 0.08$ & $1.63^{+2.34 }_{-1.19} \pm 0.17$ & 940.716 & 0.1822 & 0.41 & 0.17 & 116	& 952.6  \\ [0.2cm]
 G113.29-29.69 & NFW & pNFW & $1.86^{+0.30 }_{-0.28} \pm 0.05$ & $0.41^{+0.15 }_{-0.16} \pm 0.34$ & $0.42^{+0.83 }_{-0.41} \pm 0.61$ & $1.57^{+1.81 }_{-1.06} \pm 0.08$ & $1.84^{+2.42 }_{-1.39} \pm 0.17$ & 958.728 & 0.103 & 0.45 & 0.14 & 114	& 907.7  \\ [0.2cm]
 %G080.16+57.65 & Bur & pHer & $1.15^{+0.46 }_{-0.61} \pm 0.05$ & $2.06^{+0.85 }_{-0.96} \pm 0.34$ & $0.75^{+1.33 }_{-0.74} \pm 0.61$ & $1.60^{+1.89 }_{-1.13} \pm 0.08$ & $2.00^{+2.46 }_{-1.54} \pm 0.17$ & 347.780 & 0.08816 & 0.4299 & 634.4  \\ [0.2cm]
 G192.18+56.12 & Bur & pHer & $1.66^{+0.79 }_{-1.06} \pm 0.05$ & $2.20^{+0.63 }_{-0.73} \pm 0.34$ & $1.72^{+1.65 }_{-1.35} \pm 0.61$ & $1.35^{+2.00 }_{-0.88} \pm 0.08$ & $1.42^{+2.25 }_{-0.98} \pm 0.17$ & 834.688 & 0.1268 & 0.70 & 0.25 & 104 & 844.6  \\ [0.2cm]
 %G099.48+55.60 & Bur & pHer & $1.27^{+0.51 }_{-0.70} \pm 0.05$ & $4.01^{+4.06 }_{-2.72} \pm 0.34$ & $0.53^{+1.06 }_{-0.51} \pm 0.61$ & $1.74^{+1.88 }_{-1.27} \pm 0.08$ & $1.84^{+2.47 }_{-1.39} \pm 0.17$ & 320.995 & 0.1059 & 0.2132	& 693.1  \\ 
 \bottomrule
\end{tabular}
\end{sidewaystable*}

\begin{sidewaystable*}
\begin{tabular}{c|c|c|c|c|c|c|c|c|c|c|c|c|c}
\toprule
\hline
PSZ2 name & Mass  & $N(R)$ & $r_{200}$ & $r_{\nu}$  & $r_\text{s}$ & $\mathcal{A}_\infty$ & $\mathcal{A}_\infty$ & $-\ln P$ & $z$ & $A^2$ & $f_\text{sub}$ & $N_{200}$ & $\sigma_\text{ap,200}$\\ 
   &       &        &  [Mpc]   &  [Mpc]   & [Mpc]  & & & & & &           &   & $[\text{km/s}] $ \\ 
\midrule

 G179.09+60.12 & Bur & pNFW & $2.53^{+0.93 }_{-1.26} \pm 0.05$ & $0.67^{+0.40 }_{-0.30} \pm 0.34$ & $1.00^{+1.96 }_{-0.99} \pm 0.61$ & $1.57^{+1.92 }_{-1.10} \pm 0.08$ & $2.11^{+2.30 }_{-1.63} \pm 0.17$ & 808.742 & 0.1397 & 0.14 & 0.12 & 76 & 1195.0  \\ [0.2cm]
 G283.91+73.87 & Bur & pHer & $1.79^{+0.58 }_{-0.83} \pm 0.15$ & $1.46^{+0.65 }_{-0.42} \pm 0.14$ & $0.77^{+1.66 }_{-0.77} \pm 0.15$ & $1.56^{+1.89 }_{-1.09} \pm 0.09$ & $1.80^{+2.37 }_{-1.35} \pm 0.10$ & 671.203 & 0.08635 & 0.30 & 0.28 & 88 &  806.1  \\ [0.2cm]
 G149.39-36.84 & Eis & pHer & $1.65^{+0.53 }_{-0.66} \pm 0.15$ & $3.15^{+1.49 }_{-1.32} \pm 0.14$ & $1.42^{+1.99 }_{-1.40} \pm 0.15$ & $1.46^{+1.94 }_{-1.00} \pm 0.09$ & $2.12^{+2.35 }_{-1.63} \pm 0.10$ & 657.580 & 0.1699 & 0.44 & 0.19 & 100	& 865.8  \\ [0.2cm]
 %here
 G040.03+74.95 & NFW & pHer & $4.26^{+0.65 }_{-0.64} \pm 0.15$ & $1.64^{+0.38 }_{-0.23} \pm 0.14$ & $0.96^{+1.28 }_{-0.86} \pm 0.15$ & $1.49^{+1.76 }_{-0.95} \pm 0.09$ & $1.67^{+2.36 }_{-1.23} \pm 0.10$ & 1645.232 & 0.07067 & 4.63 & 0.39 & 241	& 2084.0  \\ [0.2cm]
 G040.03+74.95a & NFW & pHer & $2.76^{+0.49 }_{-0.45} \pm 0.12$ & 
 $1.42^{+0.41 }_{-0.37} \pm 0.29$ & 
 $1.31^{+1.89 }_{-1.21} \pm 0.60$ & 
 $1.61^{+2.65 }_{-1.15} \pm 0.11$ & 
 $1.50^{+2.58 }_{-1.05} \pm 0.08$ & 
 1186.989 & 0.07558 & 0.54 & 0.24 & 125 & 1180.0 \\ 
[0.2cm]
 G217.09+40.15 & NFW & pHer & $1.83^{+0.35 }_{-0.32} \pm 0.15$ & $1.47^{+0.49 }_{-0.39} \pm 0.14$ & $1.09^{+1.98 }_{-1.09} \pm 0.15$ & $1.40^{+1.73 }_{-0.94} \pm 0.09$ & $2.11^{+2.35 }_{-1.61} \pm 0.10$ & 1050.259 & 0.1382 & 1.42 & 0.33 & 117 & 884.8  \\ [0.2cm]
 G172.74+65.30 & NFW & pNFW & $1.57^{+0.47 }_{-0.42} \pm 0.15$ & $0.52^{+0.32 }_{-0.19} \pm 0.14$ & $1.25^{+1.96 }_{-1.24} \pm 0.15$ & $1.28^{+1.83 }_{-0.83} \pm 0.09$ & $2.05^{+2.34 }_{-1.57} \pm 0.10$ & 577.073 & 0.07529 & 0.30 & 0.20 & 74 &  720.3  \\ [0.2cm]
 G105.55+77.21 & NFW & pHer & $1.72^{+0.41 }_{-0.36} \pm 0.15$ & $1.34^{+0.58 }_{-0.42} \pm 0.14$ & $1.33^{+1.97 }_{-1.32} \pm 0.15$ & $1.31^{+1.87 }_{-0.85} \pm 0.09$ & $1.98^{+2.45 }_{-1.51} \pm 0.10$ & 700.214 & 0.07242 & 0.36 & 0.17 & 85 & 743.2  \\ [0.2cm]
 G066.68+68.44 & Bur & pHer & $1.64^{+0.35 }_{-0.29} \pm 0.15$ & $1.63^{+0.63 }_{-0.53} \pm 0.14$ & $0.36^{+0.73 }_{-0.34} \pm 0.15$ & $1.79^{+1.82 }_{-1.24} \pm 0.09$ & $1.77^{+2.41 }_{-1.32} \pm 0.10$ & 845.978 & 0.1625 & 0.15 & 0.11 & 97 &	 834.4  \\ [0.2cm]
 G050.40+31.17 & NFW & pNFW & $1.89^{+0.46 }_{-0.43} \pm 0.07$ & $0.63^{+0.37 }_{-0.24} \pm 0.52$ & $1.18^{+1.83 }_{-1.14} \pm 0.46$ & $1.33^{+1.92 }_{-0.87} \pm 0.08$ & $1.69^{+2.27 }_{-1.25} \pm 0.16$ & 837.047 & 0.1605 & 0.36 & 0.07 & 95 & 922.0  \\ [0.2cm]
 G057.78+52.32 & Eis & pHer & $1.65^{+0.29 }_{-0.27} \pm 0.07$ & $0.94^{+0.29 }_{-0.25} \pm 0.52$ & $0.32^{+0.55 }_{-0.30} \pm 0.46$ & $1.79^{+1.83 }_{-1.25} \pm 0.08$ & $2.01^{+2.39 }_{-1.55} \pm 0.16$ & 848.210 & 0.06628 & 0.23 & 0.14 & 104 & 793.8  \\ [0.2cm]
 G040.58+77.12 & NFW & pHer & $1.43^{+0.48 }_{-0.51} \pm 0.07$ & $2.52^{+1.64 }_{-1.12} \pm 0.52$ & $1.58^{+1.92 }_{-1.54} \pm 0.46$ & $1.21^{+1.77 }_{-0.75} \pm 0.08$ & $2.20^{+2.32 }_{-1.64} \pm 0.16$ & 442.840 & 0.0754 & 0.34 & 0.00 & 60	& 652.6  \\  [0.2cm]
 G028.63+50.15 & Her & pHer & $1.84^{+0.49 }_{-0.44} \pm 0.07$ & $0.73^{+0.36 }_{-0.31} \pm 0.52$ & $1.55^{+1.90 }_{-1.49} \pm 0.46$ & $1.43^{+1.94 }_{-0.97} \pm 0.08$ & $1.95^{+2.45 }_{-1.49} \pm 0.16$ & 416.306 & 0.09083 & 0.32 & 0.06 & 50 & 789.6  \\ [0.2cm]
 G186.37+37.26 & Bur & pNFW & $2.25^{+0.28 }_{-0.26} \pm 0.18$ & $0.87^{+0.24 }_{-0.14} \pm 0.14$ & $0.36^{+0.38 }_{-0.28} \pm 0.69$ & $1.79^{+1.78 }_{-1.17} \pm 0.11$ & $1.32^{+2.12 }_{-0.87} \pm 0.09$ &  2213.302 & 0.2818 & 0.48 &	0.21 & 243 & 1240.0  \\ [0.2cm]
 G228.16+75.20 & Bur & pHer & $2.54^{+0.34 }_{-0.31} \pm 0.18$ & $0.75^{+0.09 }_{-0.11} \pm 0.14$ & $0.29^{+0.24 }_{-0.19} \pm 0.69$ & $1.93^{+1.79 }_{-1.42} \pm 0.11$ & $0.95^{+1.17 }_{-0.48} \pm 0.09$ & 2705.849 & 0.5423 & 0.72 & 0.31 & 293	& 1512.0  \\ [0.2cm]
 G286.98+32.90 & Her & pNFW & $3.21^{+0.77 }_{-0.70} \pm 0.18$ & $0.31^{+0.23 }_{-0.18} \pm 0.14$ & $1.24^{+1.88 }_{-1.22} \pm 0.69$ & $1.59^{+1.92 }_{-1.12} \pm 0.11$ & $1.95^{+2.41 }_{-1.50} \pm 0.09$ & 619.176 & 0.386 & 0.16 & 0.00 & 73 & 1816.0  \\ [0.2cm]
 G028.89+60.13 & Her & pNFW & $1.70^{+0.32 }_{-0.30} \pm 0.18$ & $0.60^{+0.30 }_{-0.23} \pm 0.14$ & $1.35^{+1.85 }_{-1.22} \pm 0.69$ & $1.46^{+1.87 }_{-1.00} \pm 0.11$ & $1.82^{+2.49 }_{-1.37} \pm 0.09$ & 1001.433 & 0.1519 & 0.19 & 0.09 & 120 & 817.2  \\ [0.2cm]
 G087.03-57.37 & Bur & pNFW & $1.86^{+0.69 }_{-1.22} \pm 0.06$ & $0.72^{+0.39 }_{-0.30} \pm 0.50$ & $1.28^{+1.53 }_{-1.12} \pm 0.73$ & $1.43^{+1.87 }_{-0.97} \pm 0.09$ & $2.08^{+2.29 }_{-1.58} \pm 0.10$ & 1085.206 & 0.2761 & 0.33 & 0.32	& 97 & 1002.0  \\ [0.2cm]
 G071.63+29.78 & Bur & pHer & $1.82^{+0.89 }_{-1.12} \pm 0.06$ & $1.50^{+0.49 }_{-0.49} \pm 0.50$ & $1.97^{+1.63 }_{-1.46} \pm 0.73$ & $1.22^{+1.95 }_{-0.76} \pm 0.09$ & $1.88^{+2.35 }_{-1.34} \pm 0.10$ & 807.446 & 0.1575 & 0.41 & 0.33 & 91 & 844.3  \\ [0.2cm]
 G073.97-27.82 & NFW & pNFW & $2.48^{+0.23 }_{-0.22} \pm 0.06$ & $0.72^{+0.20 }_{-0.17} \pm 0.50$ & $0.46^{+0.46 }_{-0.34} \pm 0.73$ & $1.95^{+1.68 }_{-1.19} \pm 0.09$ & $1.04^{+1.37 }_{-0.58} \pm 0.10$ & 3368.643 & 0.2282 & 0.74 & 0.34 & 353 & 1315.0  \\ [0.2cm]
 G092.71+73.46 & NFW & pHer & $2.74^{+0.39 }_{-0.36} \pm 0.06$ & $1.95^{+0.37 }_{-0.40} \pm 0.50$ & $1.16^{+1.64 }_{-1.04} \pm 0.73$ & $1.50^{+1.85 }_{-0.97} \pm 0.09$ & $1.69^{+2.39 }_{-1.23} \pm 0.10$ & 2246.441 & 0.2317 & 0.97 & 0.18 & 261 & 1407.0  \\ [0.2cm]
 \bottomrule
 \end{tabular}
\end{sidewaystable*}
\begin{sidewaystable*}
\begin{tabular}{c|c|c|c|c|c|c|c|c|c|c|c|c|c}
\toprule
\hline
PSZ2 name & Mass  & $N(R)$ & $r_{200}$ & $r_{\nu}$ & $r_\text{s}$ & $\mathcal{A}_\infty$ & $\mathcal{A}_0$ & $-\text{ln} P$ & $z$ & $A^2$ & $f_\text{sub}$ & $N_{200}$ & $\sigma_\text{ap,200}$  \\
   &       &        &  [Mpc]   &  [Mpc]   & [Mpc]  & & & & & &           &   & $[\text{km/s}] $ \\ 
\midrule

 G187.53+21.92 & Bur & pNFW & $1.63^{+0.33 }_{-0.29} \pm 0.06$ & $0.71^{+0.31 }_{-0.23} \pm 0.50$ & $0.37^{+0.58 }_{-0.34} \pm 0.73$ & $1.59^{+1.86 }_{-1.11} \pm 0.09$ & $1.65^{+2.55 }_{-1.19} \pm 0.10$ & 1407.205 & 0.1707 & 0.53 & 0.24 & 168	& 846.2  \\ [0.2cm]
 G278.58+39.16 & NFW & pHer & $2.18^{+0.34 }_{-0.32} \pm 0.06$ & $1.75^{+0.35 }_{-0.42} \pm 0.50$ & $0.81^{+1.74 }_{-0.81} \pm 0.73$ & $1.62^{+1.78 }_{-1.10} \pm 0.09$ & $1.95^{+2.39 }_{-1.50} \pm 0.10$ & 1170.971 & 0.3063 & 0.24 & 0.26 & 130	& 1164.0  \\ [0.2cm]
 G218.81+35.51 & Bur & pHer & $3.70^{+0.57 }_{-0.61} \pm 0.06$ & $2.24^{+0.38 }_{-0.37} \pm 0.50$ & $1.85^{+1.86 }_{-1.88} \pm 0.73$ & $2.50^{+1.54 }_{-2.03} \pm 0.09$ & $3.63^{+1.50 }_{-2.20} \pm 0.10$ & 4820.696 & 0.1663 & 22.14 & 0.32 & 486	& 1852.0  \\ [0.2cm]
 %=================== a ================
  G218.81+35.51 a & Eis & pHer & $1.03^{+0.42 }_{-0.48} \pm 0.06$ & $3.42^{+0.36 }_{-0.55} \pm 0.40$ & $1.46^{+1.92 }_{-1.37} \pm 0.65 $ & $1.87^{+2.62 }_{-1.39} \pm 0.11$ & $2.09^{+2.55 }_{-1.59} \pm 0.10$ &  1357.077 & 0.1643 & 0.9179 & 0.35 & 151 & 671.7 \\
  [0.2cm]
    G218.81+35.51 b & Bur & pNFW & $1.52^{+0.46 }_{-0.43} \pm 0.07$ & 
    $0.62^{+0.23 }_{-0.25} \pm 0.22$ & 
    $0.49^{+0.80 }_{-0.44} \pm 0.72$ & $2.08^{+2.61 }_{-1.61} \pm 0.09$ & 
    $1.36^{+2.71 }_{-0.90} \pm 0.09$ & 
    843.035 & 0.1776 & 0.70 & 0.32 & 89 & 769.0 \\
  [0.2cm]
  
%========================================
 G049.32+44.37 & NFW & pHer & $1.74^{+0.38 }_{-0.37} \pm 0.06$ & $1.57^{+0.56 }_{-0.40} \pm 0.50$ & $1.71^{+1.83 }_{-1.51} \pm 0.73$ & $1.27^{+1.84 }_{-0.81} \pm 0.09$ & $1.94^{+2.46 }_{-1.44} \pm 0.10$ & 1124.578 & 0.09606 & 0.20 & 0.24 & 134 & 763.8  \\ [0.2cm]
 G046.10+27.18 & Her & pHer & $3.29^{+0.96 }_{-0.92} \pm 0.06$ & $0.57^{+0.29 }_{-0.23} \pm 0.50$ & $1.96^{+1.72 }_{-1.48} \pm 0.73$ & $1.28^{+1.93 }_{-0.81} \pm 0.09$ & $1.69^{+2.54 }_{-1.24} \pm 0.10$ & 560.821 & 0.3878 & 0.57 & 0.05 & 63	& 1598.0  \\ [0.2cm]
 G262.73-40.92 & NFW & pHer & $2.79^{+0.96 }_{-0.86} \pm 0.06$ & $0.59^{+0.22 }_{-0.23} \pm 0.50$ & $1.39^{+1.92 }_{-1.34} \pm 0.73$ & $1.24^{+2.06 }_{-0.78} \pm 0.09$ & $1.98^{+2.39 }_{-1.48} \pm 0.10$ & 530.355 & 0.4218 & 0.44 & 0.18 & 67	& 1281.0  \\ [0.2cm]
 G206.45+13.89 & Bur & pHer & $3.54^{+1.10 }_{-0.91} \pm 0.06$ & $0.98^{+0.47 }_{-0.39} \pm 0.50$ & $1.12^{+1.57 }_{-0.97} \pm 0.73$ & $1.51^{+1.96 }_{-1.06} \pm 0.09$ & $1.35^{+2.37 }_{-0.90} \pm 0.10$ & 553.161 & 0.4054 & 0.37 & 0.07 & 59	& 1496.0  \\ [0.2cm]

G083.29-31.03 & NFW & pHer & $2.44^{+0.64 }_{-0.57} \pm 0.20$ & $0.74^{+0.14 }_{-0.18} \pm 0.27$ & $1.56^{+1.90 }_{-1.48} \pm 0.51$ & $1.32^{+1.90 }_{-0.86} \pm 0.14$ & $1.86^{+2.42 }_{-1.40} \pm 0.19$ & 546.380 & 0.4102 & 0.40 & 0.00 & 21 & 1106.0 \\ [0.2cm]
 
 G201.50-27.31 & NFW & pHer & $2.35^{+0.31 }_{-0.29} \pm 0.06$ & $1.31^{+0.31 }_{-0.25} \pm 0.50$ & $1.13^{+1.55 }_{-0.98} \pm 0.73$ & $1.44^{+1.72 }_{-0.90} \pm 0.09$ & $1.65^{+2.37 }_{-1.19} \pm 0.10$ & 2281.875 & 0.539 & 0.93 & 0.31 & 241 & 1346.0  \\ [0.2cm]
G057.25-45.34 & Bur & pHer & $2.52^{+1.04 }_{-1.77} \pm 0.14$ & $4.72^{+3.45 }_{-2.61} \pm 1.22$ & $1.15^{+2.19 }_{-1.14} \pm 0.20$ & $1.65^{+1.86 }_{-1.19} \pm 0.08$ & $1.96^{+2.45 }_{-1.50} \pm 0.06$ & 479.187 & 0.3977 & 0.41 & 0.21 & 54 & 1591.0 \\[0.2cm]
 G111.61-45.71 & Her & pHer & $2.66^{+0.45 }_{-0.42} \pm 0.06$ & $0.98^{+0.21 }_{-0.16} \pm 0.50$ & $1.40^{+1.66 }_{-1.16} \pm 0.73$ & $1.37^{+1.97 }_{-0.91} \pm 0.09$ & $1.54^{+2.45 }_{-1.10} \pm 0.10$ & 2157.449 & 0.5477 & 0.76 & 0.23 & 255 & 1866.0  \\ [0.2cm]
 \bottomrule
\end{tabular}
\end{sidewaystable*}

\section{Disturbed clusters}
Here we show the projected phase space, the histogram of the line of sight velocities and the two-dimensional spatial distribution for the four clusters characterized by a value of $\delta\eta >0.8$, to highlight the evident disturbed dynamical state.
%______________________________________________________________
\begin{figure*}
   \centering
   \includegraphics[width=0.9\textwidth]{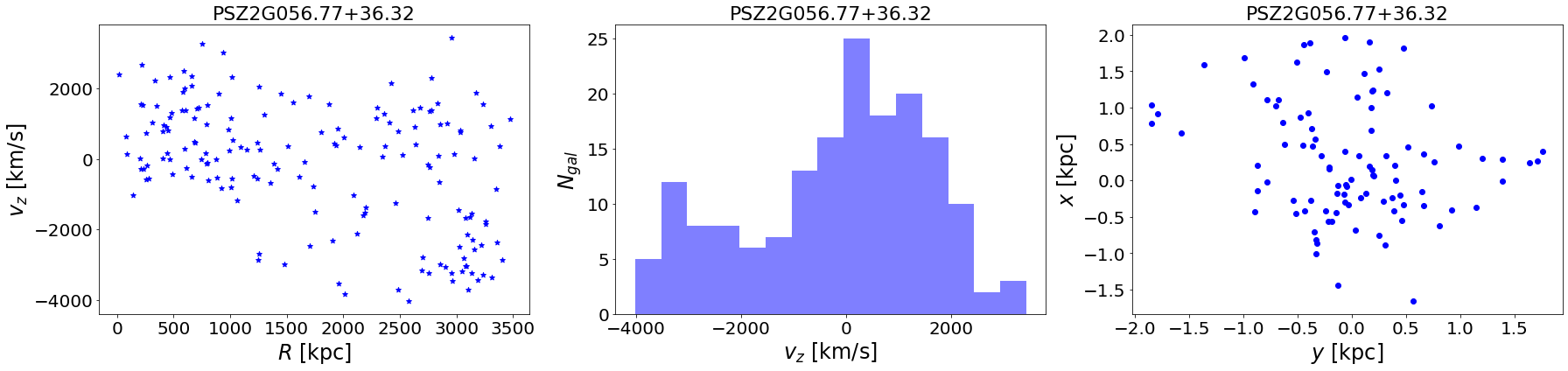}
   \includegraphics[width=0.9\textwidth]{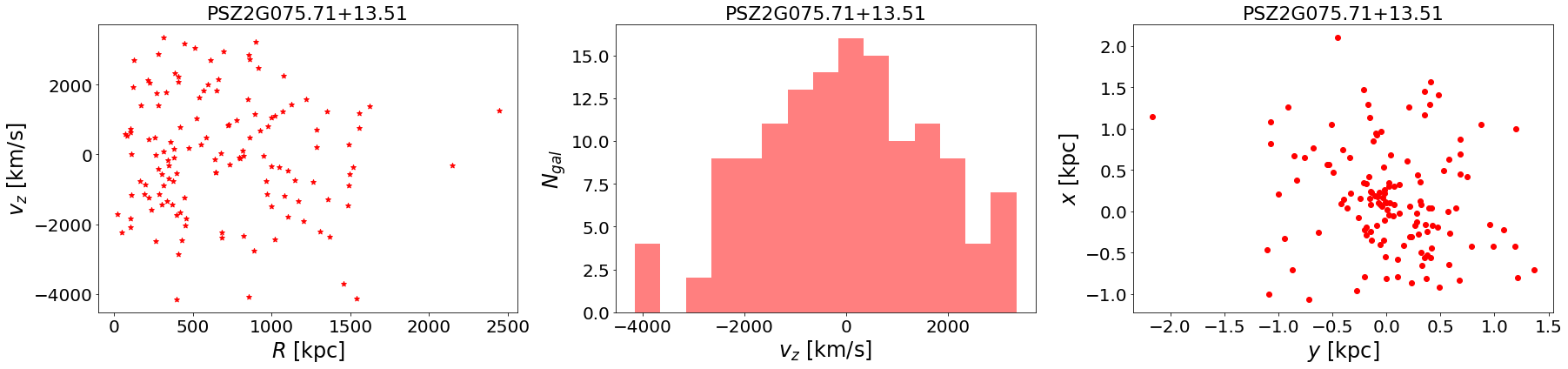}
   \includegraphics[width=0.9\textwidth]{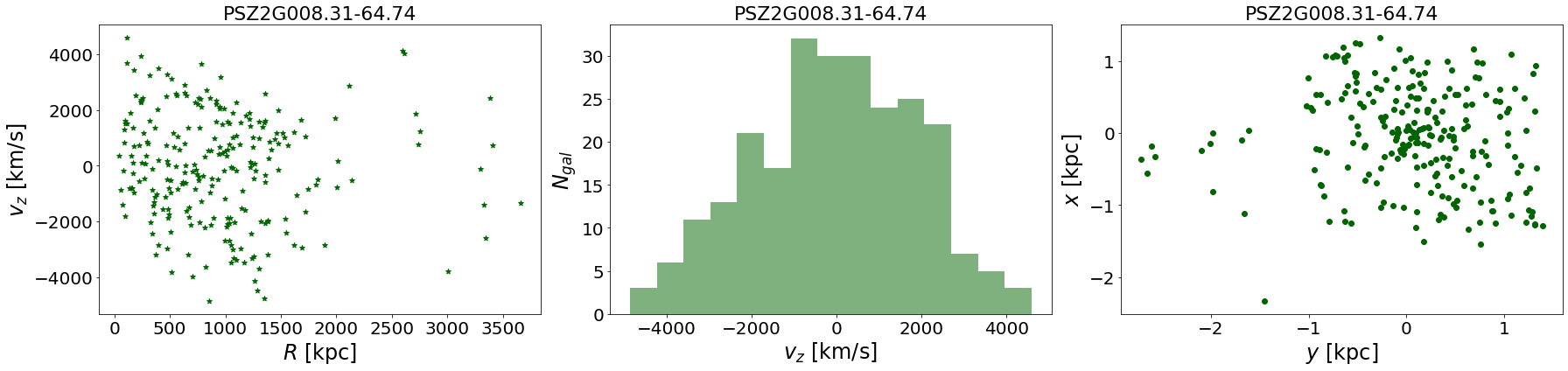}
   \includegraphics[width=0.9\textwidth]{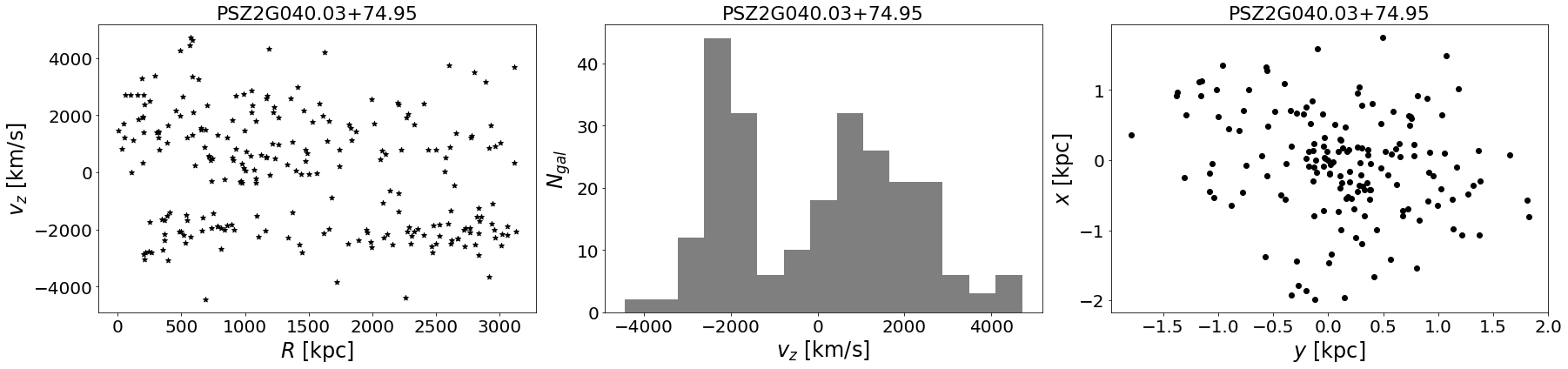}
      \caption{Each row represent one of the four clusters in the sample showing a logarithmic scatter  $\delta \eta > 0.8$ dex. Left: projected phase space. Central: LOS velocity distribution. Right: projected positions with respect to the cluster center. 
              }
         \label{fig:badpps}
   \end{figure*}
%
%______________________________________________________________

\section{Stacked projected-phase-spaces for in bins of $M^{SZ}_{500}$}
\label{app:stacking}
\begin{figure*}
   \centering
   \includegraphics[width=0.9\textwidth]{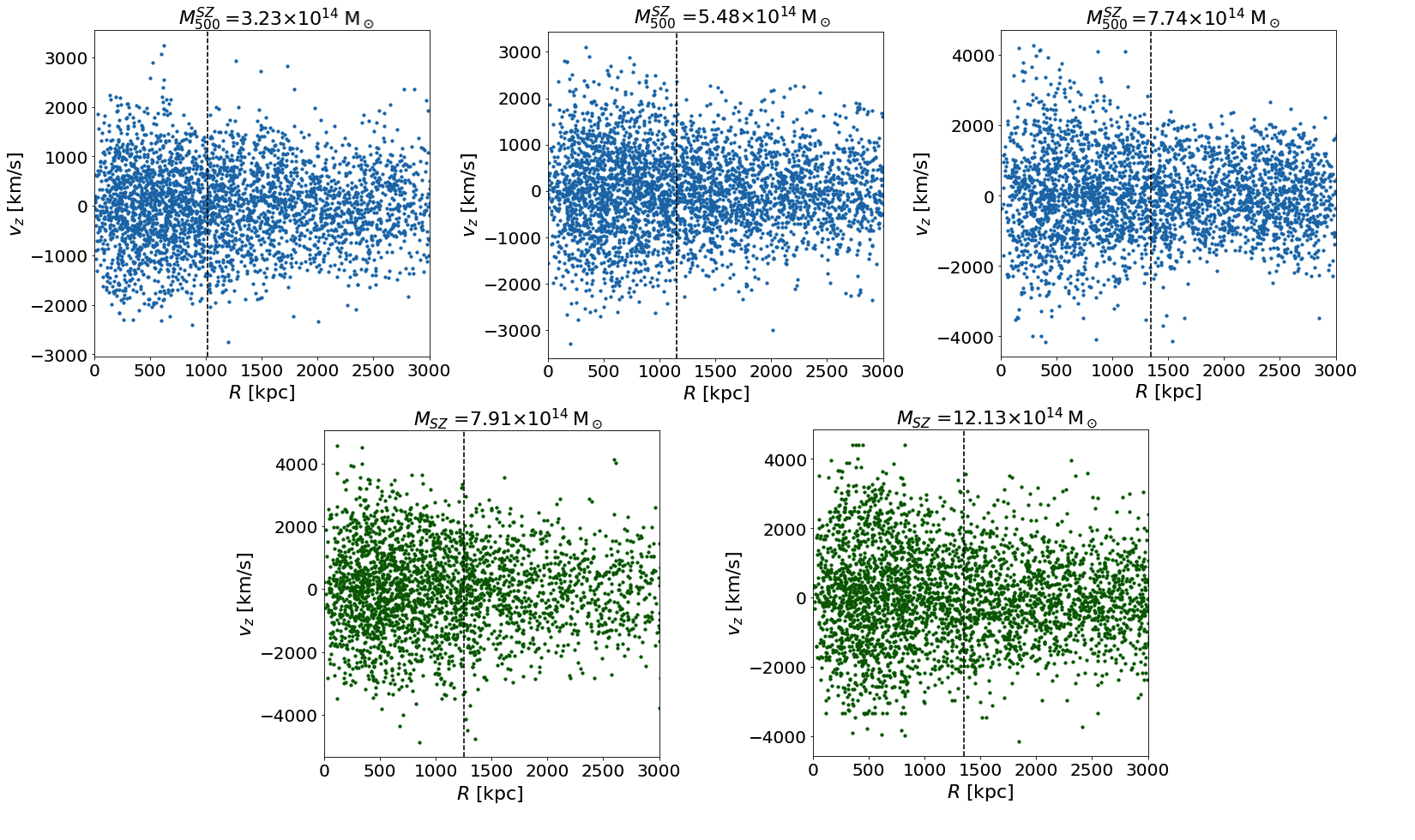}
      \caption{The projected phase space (PPS) parameters (radius $R$ and radial velocity $v$) for the SZ-stacked clusters. Top, blue points: Tier 1. Bottom, green points: Tier 2. The vertical dashed lines indicate the average $r_{500}$ of all the clusters in each stacking. 
              }
         \label{fig:SZStack}
   \end{figure*}

.
In Figure \ref{fig:SZStack} we plot the PPS of the five stacked clusters obtained by collecting member galaxies in bins of SZ masses, for Tier 1 (top plots, blue points) and Tier 2 (bottom plots, green points).
%______________________________________________________________
%
%______________________________________________________________

\section{GTC clusters data}
\label{app:data}
Here we report an extract of the member galaxies with measured spectroscopic redshifts for the three clusters observed with the OSIRIS spectrograph at the GTC. These three catalogues are organised as follow: first column corresponds to an ID label; the second and third columns indicate the right ascension and declination coordinates in J2000 epoch (RA and Dec); the fourth is the measured redshift and the last column is the errors on the redshift. The full catalogues are available upon request to the authors. 

\begin{table}
\caption{Extract of the galaxies with measured spectroscopic redshift in the field of PSZ2G083.29-31.03.}
\label{tab:fits_table}
\begin{tabular}{ccccc}
\toprule
\hline
ID & RA & Dec & z & $\delta z$ \\
 & \multicolumn{2}{c}{(deg ; J2000)} & & \\
 \midrule
1  & 337.14050 & 20.62120 & 0.41158 & 0.00009 \\
2  & 337.13602 & 20.62673 & 0.42092 & 0.00013 \\
3  & 337.15583 & 20.61639 & 0.40267 & 0.00013 \\
4  & 337.13124 & 20.59847 & 0.41477 & 0.00013 \\
5  & 337.15079 & 20.59610 & 0.42112 & 0.00009 \\
6  & 337.11054 & 20.61091 & 0.41220 & 0.00011 \\
7  & 337.10786 & 20.61638 & 0.41187 & 0.00008 \\
8  & 337.13583 & 20.59031 & 0.41003 & 0.00008 \\
9  & 337.17094 & 20.63067 & 0.41885 & 0.00013 \\
10 & 337.17612 & 20.60638 & 0.41726 & 0.00010 \\
\bottomrule
\end{tabular}
\end{table}

\begin{table}
\caption{Extract of the galaxies with measured spectroscopic redshift in the field of PSZ2G113.91-37.01.}
\label{tab:fits_table}
\begin{tabular}{ccccc}
\toprule
\hline
ID & RA & Dec & z & $\delta z$ \\
 & \multicolumn{2}{c}{(deg ; J2000)} & & \\
\midrule
1  & 4.92222 & 25.30204 & 0.36999 & 0.00007 \\
2  & 4.93030 & 25.30678 & 0.35787 & 0.00006 \\
3  & 4.93419 & 25.28389 & 0.36852 & 0.00009 \\
4  & 4.93971 & 25.31805 & 0.36313 & 0.00012 \\
5  & 4.90339 & 25.31256 & 0.36536 & 0.00008 \\
6  & 4.95634 & 25.30796 & 0.37480 & 0.00008 \\
7  & 4.90230 & 25.27815 & 0.36741 & 0.00009 \\
8  & 4.88488 & 25.28588 & 0.35860 & 0.00012 \\
9  & 4.90188 & 25.34320 & 0.37568 & 0.00008 \\
10 & 4.91762 & 25.25141 & 0.36569 & 0.00009 \\
\bottomrule
\end{tabular}
\end{table}

\begin{table}
\caption{Extract of the galaxies with measured spectroscopic redshift in the field of PSZ2G046.10+27.18.}
\label{tab:fits_table}
\begin{tabular}{ccccc}
\toprule
\hline
ID & RA & Dec & z & $\delta z$ \\
 & \multicolumn{2}{c}{(deg ; J2000)} & & \\
\midrule
1  & 262.93623 & 22.80826 & 0.38764 & 0.00037 \\
2  & 262.90941 & 22.81799 & 0.38311 & 0.00014 \\
3  & 262.91685 & 22.81963 & 0.37632 & 0.00010 \\
4  & 262.92559 & 22.82227 & 0.39349 & 0.00021 \\
5  & 262.90121 & 22.82538 & 0.37398 & 0.00027 \\
6  & 262.90976 & 22.82926 & 0.37562 & 0.00017 \\
7  & 262.90436 & 22.83141 & 0.38326 & 0.00014 \\
8  & 262.89507 & 22.83543 & 0.38148 & 0.00020 \\
9  & 262.93955 & 22.83818 & 0.37660 & 0.00033 \\
10  & 262.89412 & 22.83914 & 0.38374 & 0.00033 \\
\bottomrule
\end{tabular}
\end{table}

\end{appendix}

\end{document}